\documentclass[
reprint,
superscriptaddress,
amsmath,amssymb,
aps,
prb,
]{revtex4-1}

\usepackage{graphicx,epsf}
\usepackage{float}
\usepackage{xcolor}

\begin{document}

\title{Majorana states in prismatic core-shell nanowires}

\author{Andrei Manolescu}
\affiliation{School of Science and Engineering, Reykjavik University, 
Menntavegur 1, IS-101 Reykjavik, Iceland}
\author{Anna Sitek}
\affiliation{School of Science and Engineering, Reykjavik University, 
Menntavegur 1, IS-101 Reykjavik, Iceland}
\affiliation{Department of Theoretical Physics, Faculty of Fundamental 
Problems of Technology, Wroclaw University of Technology, Wroclaw,50-370, Poland}
\author{Javier Osca}
\affiliation{Institute of Interdisciplinary Physics and Complex Systems IFISC 
(CSIC-UIB), Palma de Mallorca, E-07122, Spain} 
\author{Lloren\c{c} Serra}
\affiliation{Institute of Interdisciplinary Physics and Complex Systems IFISC 
(CSIC-UIB), Palma de Mallorca, E-07122, Spain} 
\affiliation{Department of Physics, University of the Balearic Islands, 
Palma de Mallorca, E-07122, Spain}
\author{Vidar Gudmundsson}
\affiliation{Science Institute, University of Iceland, Dunhaga 3,
      IS-107 Reykjavik, Iceland}
\author{Tudor Dan Stanescu}
\affiliation{Department of Physics and Astronomy, West Virginia University, 
Morgantown, WV 26506, USA}

\begin{abstract}
We consider core-shell nanowires with conductive shell and insulating
core, and with polygonal cross section. We investigate the implications
of this geometry on Majorana states expected in the presence of
proximity-induced superconductivity and an external magnetic field.
A typical prismatic nanowire has a hexagonal profile, but square and
triangular shapes can also be obtained. The low-energy states are
localized at the corners of the cross section, i.\ e. along the prism
edges, and are separated by a gap from higher energy states localized
on the sides. The corner localization depends on the details of the
shell geometry, i.\ e. thickness, diameter, and sharpness of the corners.
We study systematically the low-energy spectrum of prismatic shells
using numerical methods and derive the topological phase diagram as
a function of magnetic field and chemical potential for triangular,
square, and hexagonal geometries.  A strong corner localization enhances
the stability of Majorana modes to various perturbations, including
the orbital effect of the magnetic field, whereas a weaker localization
favorizes orbital effects and reduces the critical magnetic field.
The prismatic geometry allows the Majorana zero-energy modes to be accompanied by
low-energy states, which we call pseudo Majorana, and which converge to
real Majoranas in the limit of small shell thickness.  We include the
Rashba spin-orbit coupling in a phenomenological manner, 
assuming a radial electric field across the shell.
\end{abstract}

\maketitle

\section{Introduction}

Zero-energy Majorana bound states, a concept borrowed from particle
physics, are at the center of an intense search in condensed matter
physics. Similar to the original Majorana fermions, these (quasi)particles
are identical to their antiparticles but, unlike their long-predicted
fermionic counterparts, they are characterized by non-Abelian exchange
properties. Predicted to emerge in certain types of topologically-nontrivial
quantum phases,\cite{Kitaev01,Kitaev03,Wilczek09} the zero-energy Majorana
modes benefit from the topological protection of the underlying phase and,
consequently,  represent an appealing possible platform for fault-tolerant
quantum computation.\cite{Kitaev01,Nayak08,DSarma2015, Stanescu2017}
While topological superconductors appear to be rather rare in nature,
several practical schemes for realizing topological superconductivity
and Majorana zero modes in solid state heterostructures have been
proposed in recent years.\cite{Fu08,Sau10,NadjPerge2014} The basic
physics behind these schemes has been discussed in detail in a number of
review papers,\cite{Alicea12,Stanescu13a,Beenakker13,Franz13} So far,
the most promising type of hybrid system involves  a semiconductor
nanowire with strong spin-orbit coupling proximity-coupled to a
standard s-wave superconductor and in the presence of a longitudinal
magnetic field.\cite{Lutchyn10,Oreg10} When the magnetic field exceeds
a certain critical value, the system undergoes a topological quantum
phase transition to a topologically-nontrivial superconducting phase.
In this phase, the system is predicted to host Majorana modes as
pairs of zero-energy mid-gap states localized at the two ends of the
nanowire. These Majorana zero modes are topologically-protected, in the
sense that they are robust against any local perturbation that does not
close the superconducting gap.

Extensive experimental investigations of Majorana modes in semiconductor
nanowires have been performed in the past few years. So far, the
most promising experimental signature consistent with the presence of
Majorana modes in semiconductor-superconductor hybrid structures consists
of zero-bias conductance peaks ubiquitously observed in charge transport
measurements.\cite{Mourik12,Deng12,Das12,Churchill13,Finck13,Chang2015,Krogstrup2015,Albrecht2016,Deng2016,Zhang16,Chen2016}
The conductance peak is produced by electron tunneling into the
proximitized semiconductor wire from conducting electrodes attached to its
end when a zero-energy Majorana bound state is localized in that region.
We note that this type of experiment gives only a primary indication
on the possible presence of Majorana bound states and provides no
direct evidence of a topological quantum phase transition between the
trivial to the topological superconducting phases\cite{Stanescu13a}
and no signature associated with the predicted non-Abelian exchange
properties of the Majorana zero modes.

In proximitized semiconductor nanowires, which are quasi one-dimensional
systems in symmetry class D, the topological superconducting phase that
supports the Majorana zero modes has  a ${\mathbb Z}_2$ classification. We
note that, while ideally the system is one-dimensional, real wires are, of
course, three-dimensional and exhibit a variety of non-universal phenomena
that can affect the stability of the topological phase and, implicitly,
the Majorana zero modes.  Multi-band physics\cite{Stanescu11} is an
essential aspect of this phenomenology. Another example is the orbital
effect of a magnetic field applied  parallel to the nanowire, which has
been shown\cite{Nijholt16} to be detrimental to the stability of the
the Majorana states as it reduces the energy of the low-momentum modes,
thus reducing the gap that protects the Majoranas.

The semiconductor nanowires used to realize and detect Majorana bound states are
typically grown by bottom-up methods and have a prismatic geometry,
most often with a hexagonal cross section, reflecting the underlying
crystal structure.\cite{Ihn10} With two different concentric materials
one can also obtain a prismatic core-shell heterostructure, where
the shell is conductive and the core is insulating.  
Core-shell nanowires have been recently considered for Majoranas based on semiconductor
holes, but as a 1D model only.\cite{Maier14}  Seen from a different angle, 
the polygonal
shape of the shell may be an advantage for Majorana states. In this
geometry the electronic states with the lowest energy are localized in the
corners of the shell and those with higher energies are localized on the
sides.\cite{Bertoni11} The corner localization was studied in the 1990s
in bent nanowires.\cite{Wu92,Londergan99} In our case, the cross section
of a core-shell nanowire can be seen as a (sharply) bent and closed 
channel forming a polygonal ring.

The energy separation between corner and side states is large
when the shell thickness is much smaller than the radius of the
nanowire, and when the corners are sharp, i.\ e., it increases when
the number of corners decreases.  Interestingly, apart from the
hexagonal shape, core-shell nanowires with square \cite{Hu11} or
triangular\cite{Wong11,Blomers13,Qian12,Heurlin15,Yuan15} cross section
can also be fabricated.  In particular, the gap between corner and side
states for a triangular shell of 8-10 nm and radius 50 nm can be in
the range 50-100 meV,\cite{Sitek15} i.\ e.\ larger than many detrimental
perturbations for Majorana states, including the orbital magnetic energy.
In principle, a prismatic core-shell nanowire could host several Majorana
states at each end if the corner states are completely isolated from each
other, which happens in the limit of a very narrow shell.  Therefore the
core-shell nanowires can be an experimental system with multichain ladders
discussed in recent theoretical papers.\cite{Poyhonen14,Wakatsuki14,Sedlmayr16} 
To the best of our knowledge,
the implications of the corner localization in core-shell nanowires
on the formation and stability of Majorana bound states have not been
explored yet. The main purpose of this work is to fill this gap.

The paper is organized as follows. In Section\ \ref{polyshell} we
introduce the localization of electrons in polygonal rings and in
Section\ \ref{csnws} we describe the nanowires. In Section\ \ref{soi} we
explain the SOI model.  In Section\ \ref{enor} we present energy spectra
of infinite nanowires in the normal state and in Section\ \ref{mstates}
the Bogoliubov-de Gennes spectra in the superconductor state.  In Section\
\ref{phase} we show and discuss phase diagrams for the three polygonal
geometries. 
The conclusions are collected in Section\ \ref{conc}.
Finally, in the Appendix, we present a simplified (toy) 
model of parallel chains which qualitatively reproduces the basic 
phase diagrams.

\section{\label{polyshell}Polygonal shells}

\begin{figure} 
\begin{center}
\includegraphics[width=85mm]{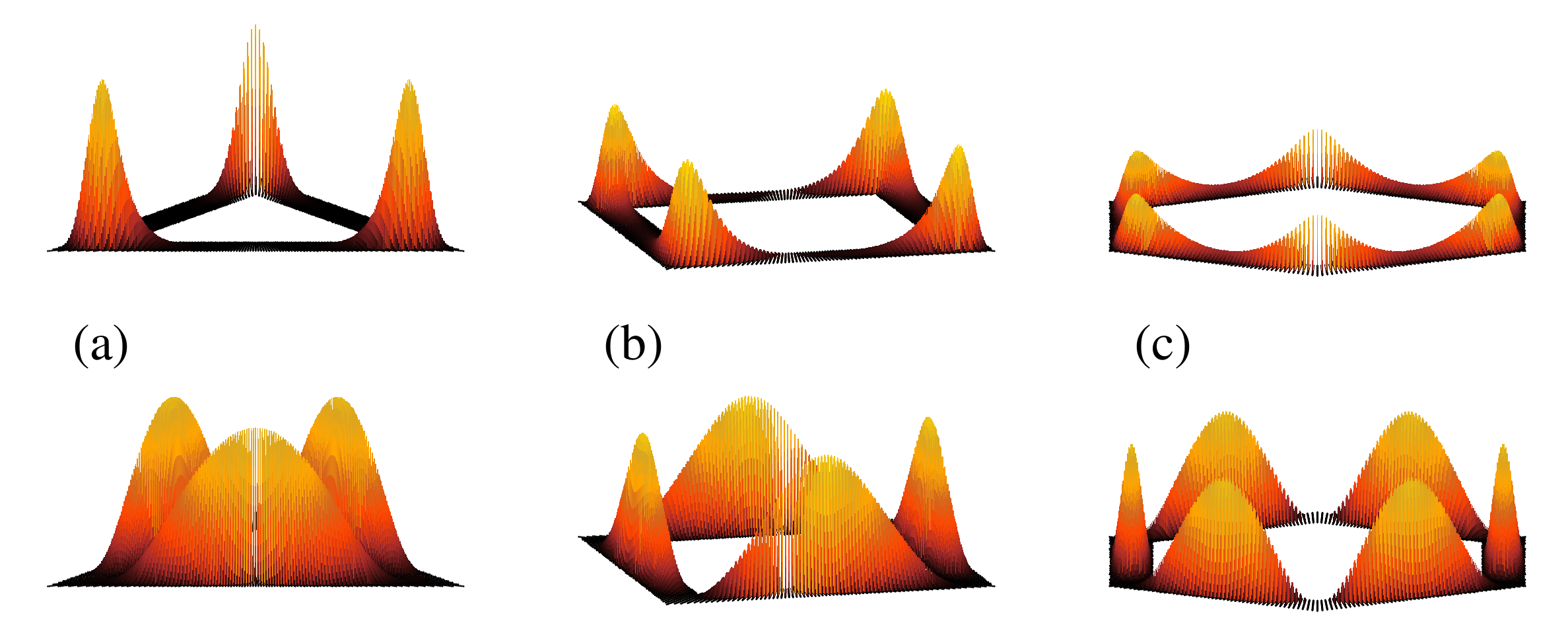}
\end{center}
\caption{Probability distributions of corner and side states of an electron
in a polygonal ring.  The upper row illustrates the corner states (including
the ground state) and the lower row the side states: (a) triangle, (b) square, 
(c) hexagon. All polygons have radius $R=50$ nm (center-to-corners) 
and side thickness $t=9$ nm.}
\label{fig:polyloc}
%
\begin{center}
\includegraphics[width=85mm]{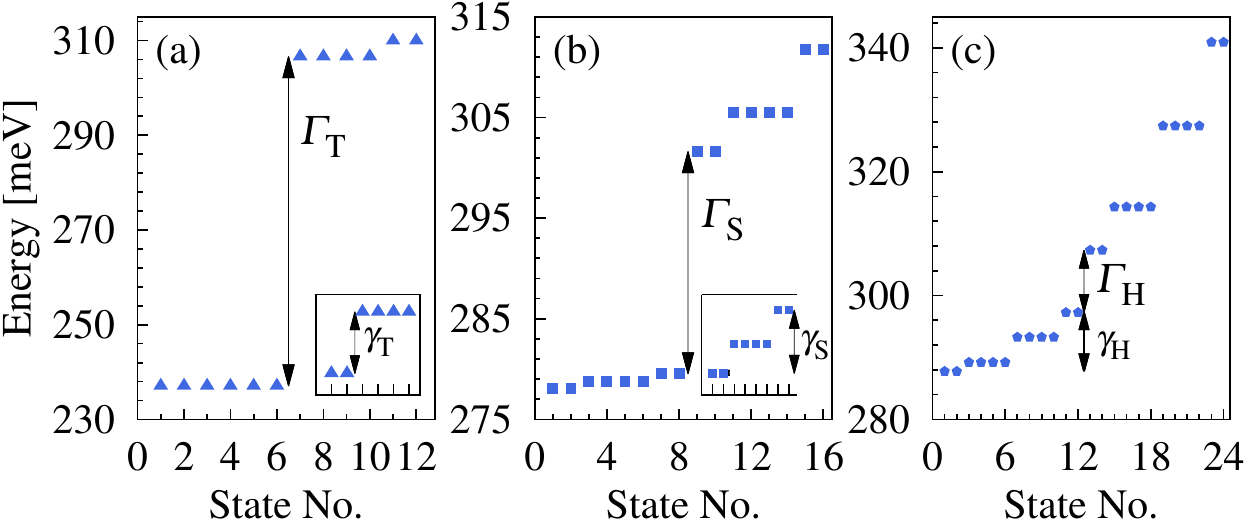}
\end{center}
\caption{Low-energy spectra for the polygonal rings shown in
Fig.\ref{fig:polyloc}.  The corner states are nearly degenerated for the
triangle (a), and have an increasing dispersion for the square (b) and
hexagon (c).  The insets show the energy span of the corner states which
is denoted by $\gamma_{\rm {T,S,H}}$.  The gap between corner and side
states, indicated as $\Gamma_{\rm {T,S,H}}$, decreases in the same order.
In the numerical calculations we used $m_{\rm eff}=0.014$, as for InSb }
\label{fig:polyene}
\end{figure}

Below we review the properties of the low-energy states in a
polygonal ring, which is the cross section of a core-shell nanowire. 
We performed the numerical diagonalization of the Hamiltonian
\begin{equation}
H_{\rm t}=-\frac{\hbar^2}{2 m_{\rm eff}} \left( \partial_x^2 +  \partial_y^2  \right) , 
\end{equation}
where $m_{\rm eff}$ is the effective electron mass in the shell material 
and the partial derivatives in the shell plane $(x,y)$
are calculated numerically within a finite-difference approximation
scheme on a grid, with Dirichlet boundary conditions at the
edges of the polygons.\cite{Sitek15,Sitek16}  To reach convergence the
grid included several thousands of points.  In Fig.\ \ref{fig:polyloc}
we show typical probability distributions of corner and side states for a
symmetric triangle, square, and hexagon, all with the same side thickness
$t=9$ nm and circumference radius $R=50$ nm, {for InSb parameters (see Sec. III)}.  
For each polygon there are
$2N$ corner states, where $N$ is the number of corners, followed on the
energy scale by $2N$ side states, the counts including the spin.

For this aspect ratio, ${\rm AR}=t/R=0.18$, the corner states of
the triangle consist in nearly isolated peaks and quasi-degenerate
six energy levels, with a small dispersion $\gamma_{\rm T}=0.02$ meV,
separated  from the side states by a gap $\Gamma_{\rm T}=70$ meV, Fig.\
\ref{fig:polyene}(a).  Next, for the square, the corner localization
softens a little bit, the eight corner states have a broader dispersion,
$\gamma_{\rm S}=1.4$ meV, and the energy separation from the side states
decreases to $\Gamma_{\rm S}=22$ meV, Fig.\ \ref{fig:polyene}(b). Further,
for the hexagon, the corner peaks drop more and have tails onto
the polygon sides, the corresponding twelve corner states have a
considerable dispersion, $\gamma_{\rm H}=8$ meV, comparable to the
interval between corner and side states $\Gamma_{\rm H}=10$ meV, Fig.\
\ref{fig:polyene}(c).  Higher energy states (not shown in the figures)
have increasingly spread localization.
Apart of the spin degeneracy, the symmetries of the polygons lead to
orbital double degeneracies, such that the degeneracy sequences are (24,
42,...), (242, 242,...), and (2442, 2442,...), for triangle, square,
and hexagon, respectively.\cite{Sitek15}  By decreasing the AR of the 
polygons the corner localization becomes stronger, the probability
distribution converges to totally isolated peaks for each polygon, and
the energy separation between the highest corner state and the lowest
side state considerably increases.\cite{Sitek16}

\section{\label{csnws}Models of core-shell nanowires}

We build the Hamiltonian of the nanowire, $H_{\rm w}$, from several terms,
\begin{equation} 
H_{\rm w}=H_{\rm t}+H_{\rm \ell}+H_{\rm Z}+H_{\rm SOI} .
\label{eq:Hw}
\end{equation} 
We consider a magnetic field along the nanowire, i.\ e.\ along the
$z$ axis, ${\bf B}=(0,0,B)$, and incorporate it in the transverse Hamiltonian
$H_{\rm t}$ via the canonical momentum $p_{\varphi}+eA_{\varphi}$, where 
$A_{\varphi}=Br/2\hat{\boldsymbol\varphi}$ is the vector potential in
polar coordinates $(\varphi,r)$ transversal to the  nanowire, and 
$p_{\varphi}=(-i\hbar/r)\partial_{\varphi}$.

The second term of Eq.\ (\ref{eq:Hw}) corresponds to the longitudinal 
degree of freedom, $H_{\rm \ell}=p_z^2/(2m_{\rm eff})$.  For an infinite nanowire its 
eigenstates are the plane waves $| k \rangle = \exp (ikz) / \sqrt{L}$, where $k$ is 
the wave vector, $L\to \infty$ being the nanowire length.  For a finite length $L$
we assume hard wall boundaries at $z=\pm L/2$ and the eigenstates become 
$| n \rangle =  \sqrt{2/L} \sin \left[ n(z/L+1/2)\pi \right]$, with $n=1,2,3,...$.

Next, the term $H_{\rm Z}=-g_{\rm eff}\mu_{\rm B} \sigma B$ is the spin Zeeman term, 
with $g_{\rm eff}$ the effective g-factor in the shell material, $\mu_{\rm B}$ Bohr's
magneton and $\sigma=\pm 1$ the spin quantum number, and finally $H_{\rm SOI}$ is the 
spin-orbit term, which will be discussed in the next section.

In the calculations we use material parameter values corresponding to 
(bulk) InSb, which is one of the most interesting semiconductors for Majorana 
detection due to the large g-factor and strong SOI.
We used $m_{\rm eff}=0.014$ and $g_{\rm eff}=-51.6$,
i.\ e. the values known for the bulk material, although they may be 
different for our thin shell structures. To find
them properly one needs to perform ab initio calculations starting with
the proper atomic structure in the prismatic geometry.  This is a complex
problem which is beyond the scope of our present work.  
We thus compute the electronic states at a mesoscopic scale,
i.\ e. by averaging over many unit cells of the atomic lattice, such
that the dominant effects are due to the geometry. This approach has
been able to describe the relevant physics related to edge or side
localization.\cite{Bertoni11,Wong11} 
Still, the precise quantitative
validity for core-shell nanowires of simplifying approximations such as
the use of piecewise constant potentials and material parameters is
something to be confronted with more refined models in the future. The
intention of our present work is to predict several scenarios of
Majorana physics in shells of prismatic geometry, using the bulk
parameters as test values, i.\ e. mid- way between a qualitative and a
quantitative style, and to support each scenario with plausible
examples. The qualitative agreement with a toy model (discussed in
Appendix) indicates a rather robust physical behavior of the phase
diagrams discussed below.

\section{\label{soi}Spin-orbit interaction}

\begin{figure} [t]
\begin{center}
\includegraphics[width=75mm]{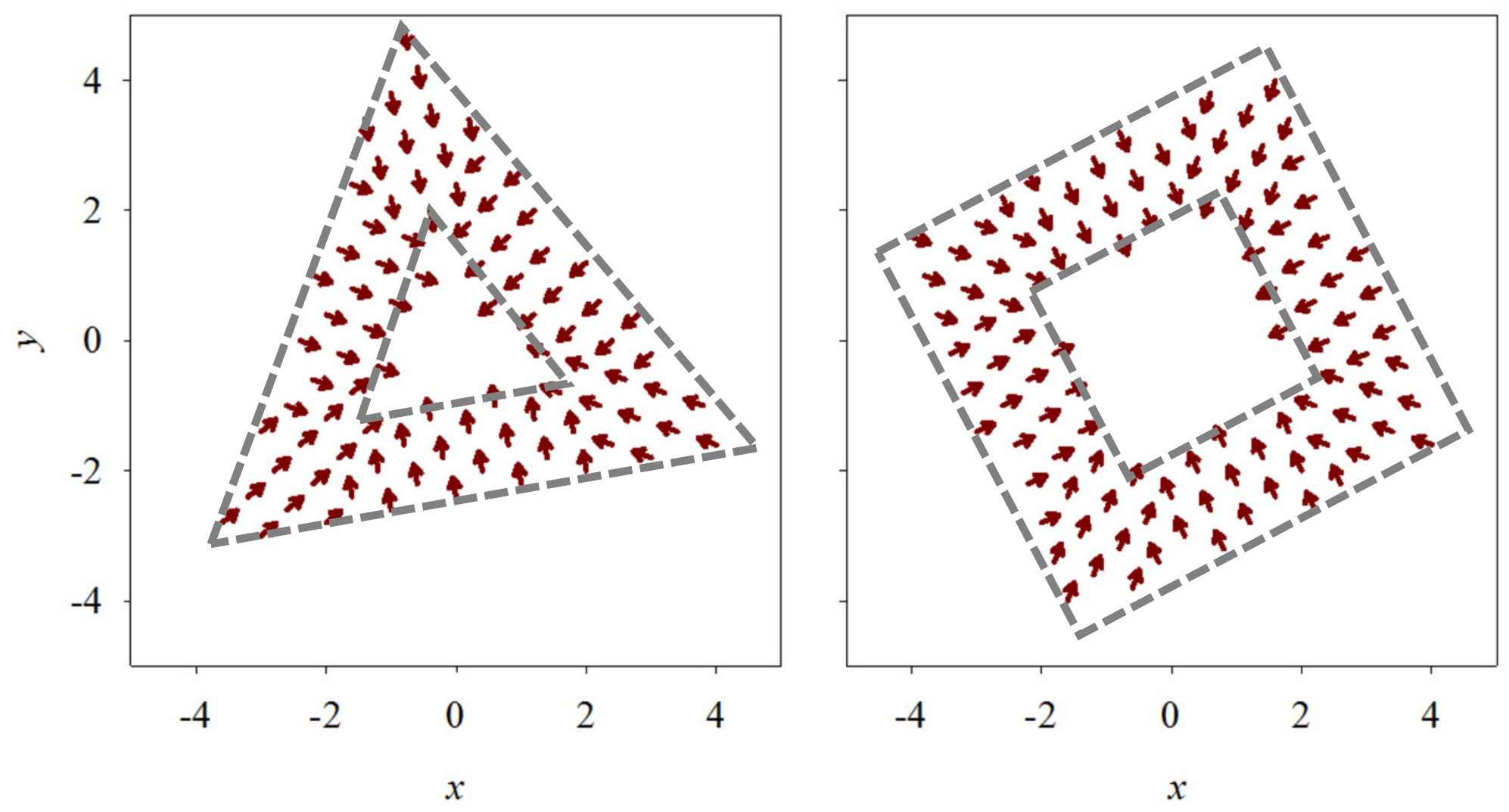}
\end{center}
\vspace{-5mm}
\caption{
The effective Rashba electric field for a square and a triangle in 
the prismatic SOI model.  The corner areas span angular intervals of $40^o$. 
}
\label{fig:Rfield}
\end{figure}

In order to obtain Majorana states we need a Rashba-type SOI model for
the prismatic geometry.  
In heterostructures where materials with different work functions
are placed next to each other interface electric fields are generated. 
For a planar 2D electron system created
in a semiconductor heterostructure the origin of SOI is the effective (net) 
electric field, normal to the interface, associated with an asymmetric
confinement. 
For a core-shell heterostructure we assume 
similar intrinsic fields, present at the core-shell and/or at the shell-vacuum interfaces,
perpendicular to the lateral surfaces of the prismatic shell, and thus changing
direction at the corners, as illustrated in Fig.\ \ref{fig:Rfield}.
In core-shell nanowires there is a geometric asymmetry 
in the radial direction, inwards vs. outwards, that lends plausibility to 
an effective radial field. In principle such a field could 
also be obtained or controlled with gates.
In cylindrical coordinates the field has radial and azimuthal components,
${\bf E}=(E_{\varphi},E_r,0)$.  The SOI Hamiltonian can be calculated as
\cite{Winkler04,Bringer11}
\begin{equation}
H_{\rm SOI}=\frac{\lambda}{\hbar}{\boldsymbol \sigma}\left( {\bf p} \times {\bf E} \right) ,
\label{eq:SOIpri}
\end{equation}
where $\lambda$ is the SOI coupling constant and ${\boldsymbol \sigma}$
are the Pauli matrices.  We implement both $E_{\varphi}$ and $E_r$
as functions of $\varphi$ with two independent strength parameters.
This is a phenomenological model
aimed at capturing the basic SOI effect dictated by the geometry of the
heterostructures, which is necessary for the realization of topological
superconductivity and Majorana bound states.
In reality the Rashba coupling, and also the g-factor, are expected
to depend on the shell thickness and, possibly, on the radius of the
wire. To evaluate them rigorously a computationally involved approach is  
necessary, for example based 
on a multiband $k\cdot p$ model\cite{Winkler04}, taking into account
the core-shell materials and geometry. This is an important task, but 
still, outside our present focus.

In particular, for a cylindrical shell, following the same line of arguments,
the effective interface field should be radial and constant, i.\ e. ${\bf E}=(0,E,0)$. 
In this case Eq.\ (\ref{eq:SOIpri}) gives
\begin{equation}
H_{\rm SOI}=\frac{\alpha}{\hbar}(\sigma_{\varphi}p_z-\sigma_z p_{\varphi}) ,
\label{eq:SOIrad}
\end{equation}
where $\alpha=\lambda E$.  We shall call (\ref{eq:SOIrad}) and
(\ref{eq:SOIpri}) the cylindrical and prismatic SOI models, respectively.
The cylindrical model can also be seen as a straight forward transformation
of the standard two-dimensional model where the planar Cartesian coordinates
are replaced by cylindrical coordinates.\cite{Bringer11,Manolescu13}

For the prismatic nanowires, as long as we consider only corner states in
the shell, the SOI is restricted to the corner areas, where we can assume
$E_{\varphi}\approx 0$ and a constant $E_r$.  Therefore, by neglecting
the presence of the electrons on the sides of the polygonal shell, where
the wave functions exponentially vanish (Fig.\ \ref{fig:polyloc}), the
cylindrical model should still be reasonable and qualitatively correct.
In the numerical calculations we included the SOI using the
cylindrical model, with $\alpha=50\ {\rm meV \ nm}$ (as for InSb).
We also tested the energy spectra with the prismatic SOI model (Fig.\
\ref{fig:Rfield}), and we obtained similar results. In this context
it is worth mentioning that the cylindrical SOI model cannot lead to
Majorana states for a wire with circular symmetry.  \cite{Lim13}  As we
shall see, this is no longer true for the prismatic geometry.

\section{\label{enor}Energy spectra of nanowires in the normal state}

In order to obtain the eigenstates of a nanowire we first diagonalize the
Hamiltonian of the polygonal cross section, $H_t$, on the transverse grid.
Then, we combine the $N$ corner states, which are the lowest-energy states
of $H_t$, $|a\rangle$, $a=1,2,...,N$, with the longitudinal modes,
which are the eigenstates of $H_{\ell}$, $|k\rangle$ or $|n\rangle$,
for the infinite or finite length, respectively.  By adding the spin we
form a basis set in the total Hilbert space, $|g\rangle=|ak\sigma\rangle$
for the infinite wire and $|g\rangle=|an\sigma\rangle$ for the wire of
finite length.  We use these bases to calculate the matrix elements of
the total Hamiltonian, $\langle g|H_{\rm w}|g'\rangle$.  The matrices
are diagonalized numerically to give the eigenstates of the two models.
For the infinite nanowire $H_{\rm w}$ is already diagonal in $k$, and we
calculate the eigenstates within the subspace $|a\sigma\rangle$ for an
array of $k$ values.  For the case with finite length we diagonalize a
single, but larger matrix including the longitudinal modes $|n\rangle$.
The basis set is truncated appropriately for convergence.  We used
only the corner states as transverse modes $|a\rangle$ because the
SOI and other perturbations are too weak to mix them significantly 
with the side states.  By including the side states the largest correction, 
expected for the highest corner states in the hexagonal shell, 
is at most 1 meV.
For the nanowires with finite length we included all longitudinal modes
up to $n=200$.

\begin{figure} [t]
\begin{center}
\includegraphics[width=85mm]{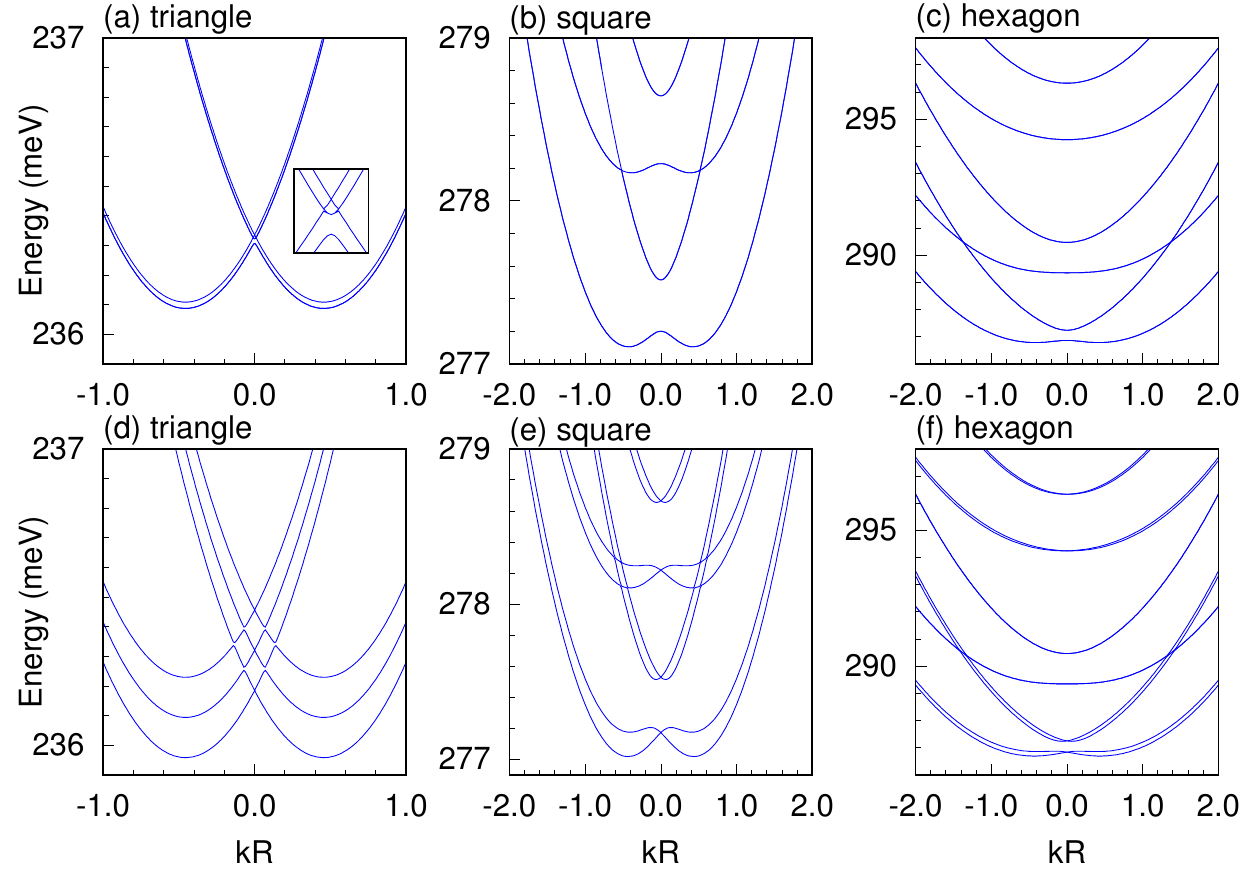}
\end{center}
\vspace{-5mm}
\caption{Energy dispersions for the corner states of infinite nanowires
with polygonal shells including SOI within the cylindrical model.  
The top row shows $2N$ bands for the three polygonal shapes:  
(a) triangular, $N=3$, (b) square, $N=4$, (c) hexagonal, $N=6$.  
{Due to the geometric symmetries all bands are degenerate, except
one pair for the triangle, crossing at $k=0$, as shown in
the zoomed inset.}
The bottom plots show the results with a transverse electric field
of 0.22 mV$/R$ which perturbs the symmetry of each polygon,
{lifting all spin degeneracies at $k\neq 0$.} 
Here the radius $R=50$ nm and side thickness $t=9$ nm.  The
material parameters are as for InSb.
}
\label{fig:Enor}
\end{figure}

In Fig.\ \ref{fig:Enor} we display energy eigenvalues of $H_{\rm
w}$ for infinite nanowires with the three polygonal cross sections
discussed: triangular, square, and hexagonal.  In the absence of SOI all
energies (or bands) are parabolic functions of $k$. {With SOI they
remain even functions, but non-monotonic for $k>0$ or $k<0$. For symmetric
polygons, i.\ e.  with equal corner angles, the energy bands are shown in
Fig.\ \ref{fig:Enor}(a-c).  In these cases the eigenstates of $H_{\rm
w}$ appear pairwise degenerate for each fixed $k$, although the spin
is not conserved.  For the square and the hexagon four and six bands, 
respectively, can be observed in the figures, which in reality are 
eight and twelve, i.\ e. the number of corners times spin.
In the triangular case, due to the absence of inversion symmetry, 
one degeneracy is lifted for any $k\neq 0$, and only two degenerate
bands are obtained, and thus four energy curves are seen in Fig.\
\ref{fig:Enor}(a).  At $k=0$ the nondegenerate bands have the familiar
crossing (where SOI vanishes), and this is the only exact level crossing
point in Fig.\ \ref{fig:Enor}.}

For the purpose of our work the strictly symmetric polygons should be
regarded only as mathematical models.  The prismatic nanowires grown
in labs are naturally not perfectly symmetric even if the polygonal
cross section is a result of a specific lattice structure.  In addition,
in order to detect Majorana states, the experimentalists use gates and
contacts that are expected to break the polygonal symmetry.  Therefore,
in our model we perturb the polygonal symmetries by imposing an electric
field along one side of each type of polygon, which generates a voltage
of 0.22 mV across the length of one radius.  This perturbation lifts
the orbital degeneracies of the transversal modes of the nanowires
(shown earlier in Fig.\ \ref{fig:polyene}), and all energy bands
for each polygonal shape are now spin split, as seen in Fig.\
\ref{fig:Enor}(d-f). {Still, the splitting decreases for large
polygon angles and it is not visible for all hexagon bands in (f).}

\section{\label{mstates}Majorana states}

The Majorana states are obtained using the Bogoliubov -- de Gennes Hamiltonian (BdG),
$H_{\rm BdG}$, which we obtain with the matrix elements of $H_{\rm w}$ and with the 
isospin quantum number $\tau=\pm 1$.  For the infinite wire they can be written 
as
\begin{equation}
\begin{split}
& \langle a \sigma \tau | H_{\rm BdG} (k) | a' \sigma' \tau\rangle = 
\tau \operatorname{Re}\langle a\sigma\tau|H_{\rm w}(\tau k)|a'\sigma'\tau\rangle \\
& +i\operatorname{Im} \langle a\sigma\tau|H_{\rm w}(\tau k)|a'\sigma'\tau\rangle 
- \tau \mu\delta_{aa'}\delta_{\sigma\sigma'} ,
\end{split}
\label{eq:infwdiag}
\end{equation}
\begin{equation}
\langle a \sigma \tau | H_{\rm BdG} (k) | a' \sigma' \tau'\rangle = 
\tau \sigma\delta_{\sigma,-\sigma'}\delta_{aa'}\Delta , \ \tau\neq\tau' .
\label{eq:infoffdiag}
\end{equation}
Eqs.\ (\ref{eq:infwdiag}-\ref{eq:infoffdiag}) define the diagonal and off-diagonal
elements in the isospin space, respectively. $\mu$ is the chemical potential, 
$2\Delta$ is the superconductivity gap, and $\delta$ denotes the Kronecker symbol.
Here the wave vector $k$ is included in the Hamiltonian because it 
behaves like  a parameter. For the nanowire of finite length these equations 
are replaced by
\begin{equation}
\begin{split}
& \langle g \tau | H_{\rm BdG} | g' \tau\rangle = 
\tau \operatorname{Re}\langle g\tau|H_{\rm w}|g'\tau\rangle \\
& +i\operatorname{Im} \langle g |H_{\rm w}|g'\tau\rangle 
- \tau \mu\delta_{gg'} ,
\end{split}
\label{eq:finwdiag}
\end{equation}
\begin{equation}
\langle g \tau | H_{\rm BdG} | g' \tau'\rangle = 
\tau \sigma\delta_{\sigma,-\sigma'}\delta_{aa'}\delta_{nn'}\Delta , \ \tau\neq\tau' .
\label{eq:finoffdiag}
\end{equation}
where we used the previous notation $|g\rangle=|a n \sigma\rangle$ for the 
basis states of the nanowire in the normal state.

\begin{figure} [t]
\begin{center}
\includegraphics[width=85mm]{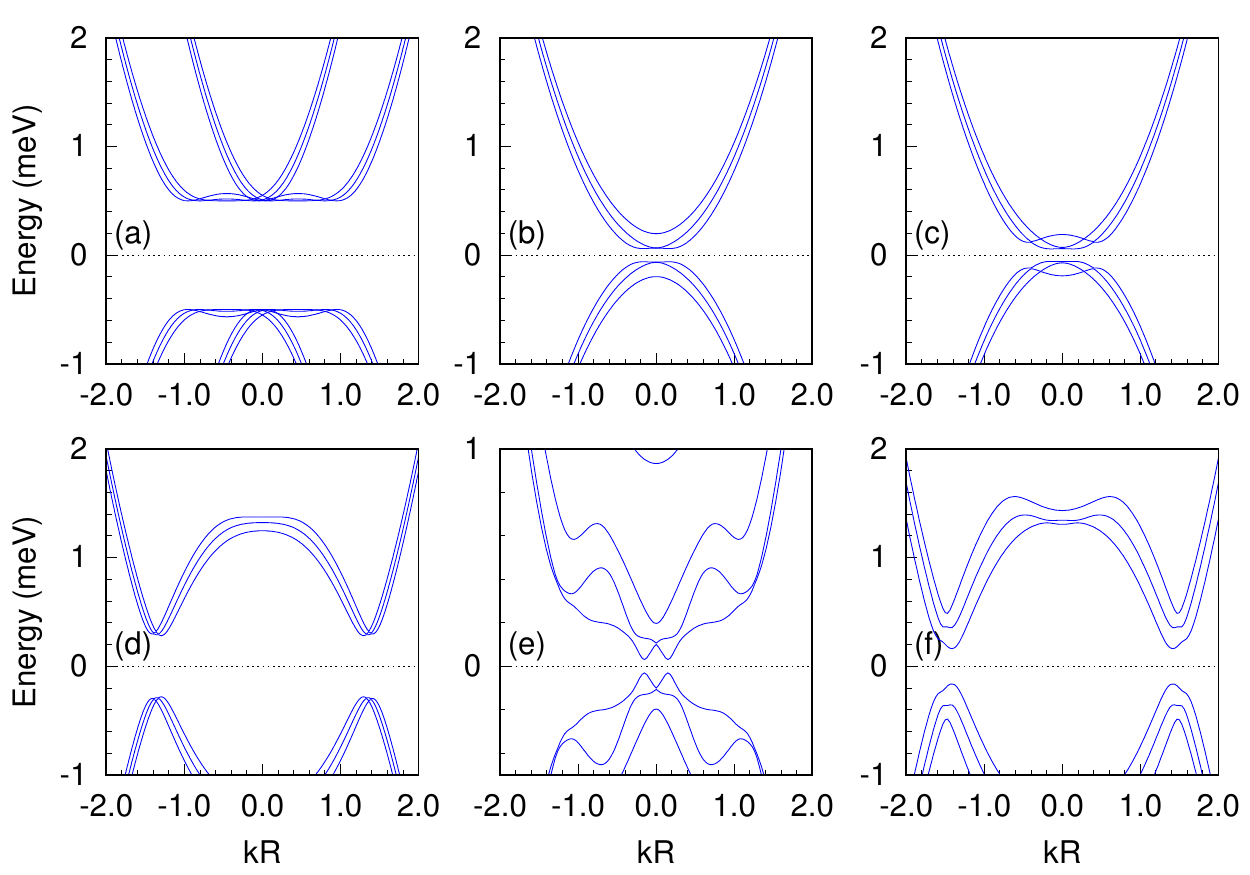}
\end{center}

\vspace{-5mm}
\caption{{\bf Triangular shell.} 
BdG eigenenergies vs. the wave vector for a triangular shell
of infinite length.  The superconductor gap
closes and reopens depending on the values of the chemical potential $\mu$
and magnetic field $B$. The nanowire radius $R=50$ nm is fixed. The
shell thickness is $t=9$ nm, i.\ e. ${\rm AR}=0.18$ for the cases (a)-(d).
Other parameters:
(a) $\mu=236.2$ meV,\ $B=0$, 
(b) $\mu=234.6$ meV,\ $B=1.32$ T, 
(c) $\mu=234.7$ meV,\ $B=1.32$ T, 
(d) $\mu=236.2$ meV,\ $B=1.32$ T.
Next, $t=12.5$ nm, i.\ e. ${\rm AR}=0.25$, for
(e) $\mu=126.7$ meV,\ $B=0.34$ T, and 
(f) $\mu=126.7$ meV,\ $B=1.32$ T. 
}
\label{fig:EBdGk_tri}
\end{figure}

\subsection{BdG spectra for nanowires of infinite length}

In Fig.\ \ref{fig:EBdGk_tri} we show several examples of BdG spectra for
a triangular nanowire of infinite length.  We use the bare superconductor
energy parameter $\Delta=0.5$ meV.  Only the transversal corner states
are considered in the calculations, the mixing with the side states
being completely negligible for the chemical potentials and magnetic
fields used.  The spectra are selected for the further discussion on
the phase diagrams with several representative values of the chemical
potential and magnetic field $B$.  The triangular symmetry is broken
by the small electric bias discussed in the previous section.  In the
first example, Fig.\ \ref{fig:EBdGk_tri}(a), the BdG states are obtained
with $\mu$ near the intersection of the six energy curves seen in Fig.\
\ref{fig:Enor}(d) and $B=0$. This case corresponds to the trivial phase
of the proximity-induced superconductivity in the nanowire.  The spectra
are particle-hole symmetric, but for clarity we display a larger interval
for particle than for hole states.

By varying the chemical potential and/or increasing the magnetic field,
and thus the Zeeman energy, the gap closes first at $k=0$, and then it
opens again, and the systems enters into the topological phase.  
In Figs.\ \ref{fig:EBdGk_tri}(b,c) we show
two situations with $B=1.32$ T, and a very small difference between the
chemical potentials, $\mu=234.6 \ {\rm and} \ 234.7$ meV, respectively,
such that one can see a small gap at $k=0$ reopening due to the
repulsion of different corner states.  In Fig.\ \ref{fig:EBdGk_tri}(d),
with the same magnetic field, and $\mu=236.2$ meV, the gap at $k=0$ is
largely open.  In these cases Majorana states are expected, located at
the ends of the nanowire, and with zero energy, which we shall observe
with the nanowire model of finite length.  Notice the smaller gaps at
finite $k$ in Figs.\ \ref{fig:EBdGk_tri}(d,e,f).  They indicate orbital
effects of the magnetic field which may possibly forbid the formation
of Majorana states if such gaps become very small (much smaller than
$2\Delta$).\cite{Nijholt16} By increasing the thickness of the shell, from
9 to 12.5 nm, the energy separation between the corner states, and hence
between the bands shown grows, as seen in Figs.\ \ref{fig:EBdGk_tri}(e,f).
At the same time the corner localization weakens and the orbital effects
increase, so the Majorana suppression at large magnetic fields becomes
more likely, Figs.\ \ref{fig:EBdGk_tri}(e).

\begin{figure} [t]
\begin{center}
\includegraphics[width=85mm]{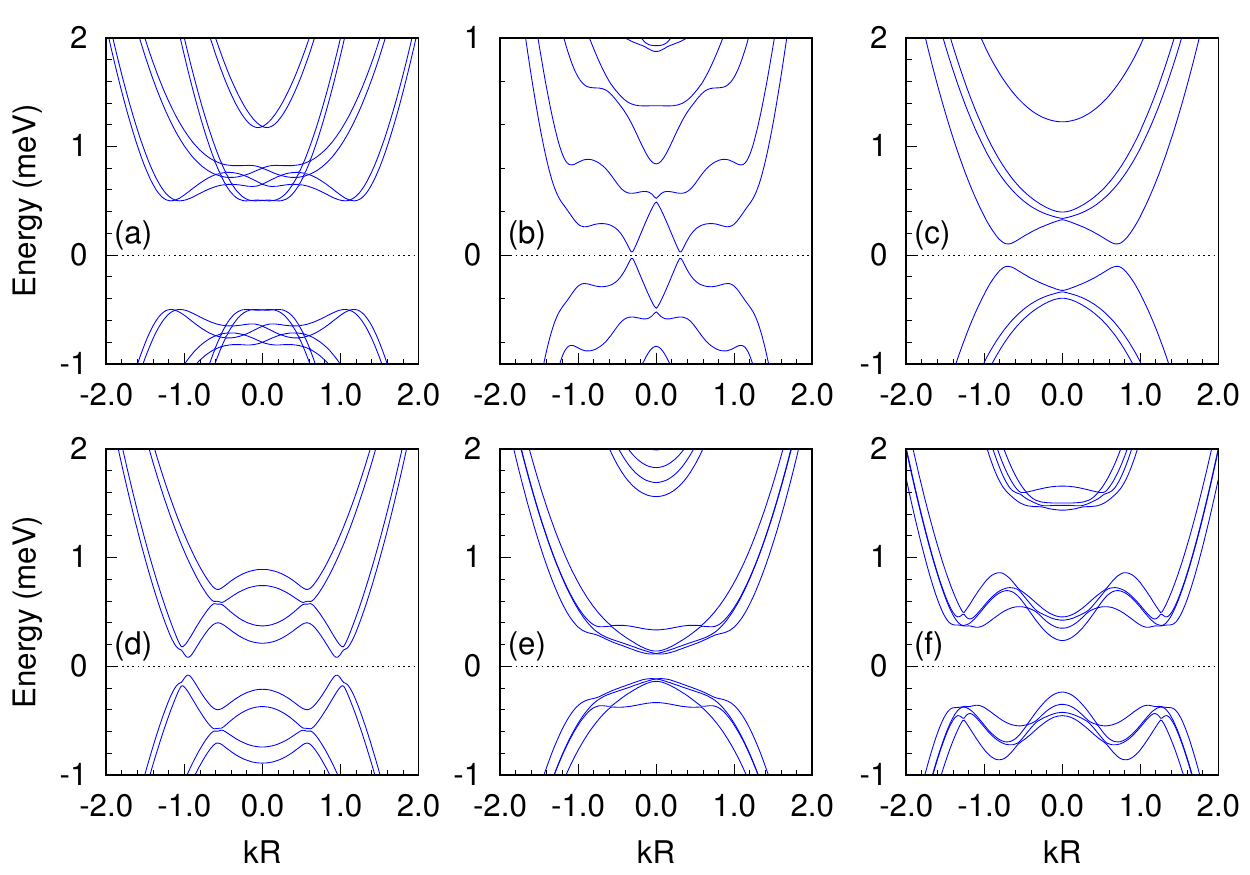}
\end{center}
\vspace{-5mm}
\caption{{\bf Square shell.}
{BdG energy spectra similar to those shown in Fig.~\ref{fig:EBdGk_tri}, 
for a nanowire of radius $R=50$ nm.} The 
shell thickness is $t=9$ nm, i.\ e. ${\rm AR}=0.18$ for the cases
(a) $\mu=277.6$ meV,\ $B=0$,  
(b) $\mu=277.6$ meV,\ $B=0.33$ T, 
(c) $\mu=276.5$ meV,\ $B=0.79$ T,
(d) $\mu=276.5$ meV,\ $B=1.32$ T. 
Next, $t=8$ nm,  i.\ e. ${\rm AR}=0.16$ for 
(e) $\mu=346.4$ meV, \ $B=0.66$ T, and 
(f) $\mu=347.3$ meV,\ $B=0.66$ T. 
}
\label{fig:EBdGk_squ}
\end{figure}

In Fig.\ \ref{fig:EBdGk_squ}, we show several cases with the square
geometry, with four corner states (not counting the spin), which lead
to more complicated spectra.  The topologically trivial phase is shown
in Fig.\ \ref{fig:EBdGk_squ}(a), for a shell thickness of 9 nm.  Fig.\
\ref{fig:EBdGk_squ}(b), with nearly closed gaps at finite $k$, and Fig.\
\ref{fig:EBdGk_squ}(c), with a larger gap at a larger $k$, correspond
to a topological phase.  The square geometry implies a larger energy
separation between the corner states compared to the triangular geometry
with the same AR, and thus the orbital effect may be more detrimental
for the square than for the triangular shell.  By slightly reducing the
shell thickness from 9 to 8 nm we can obtain more robust topological
phases for the square geometry, Fig.\ \ref{fig:EBdGk_squ}(d,e,f).

\begin{figure} [t]
\begin{center}
\includegraphics[width=85mm]{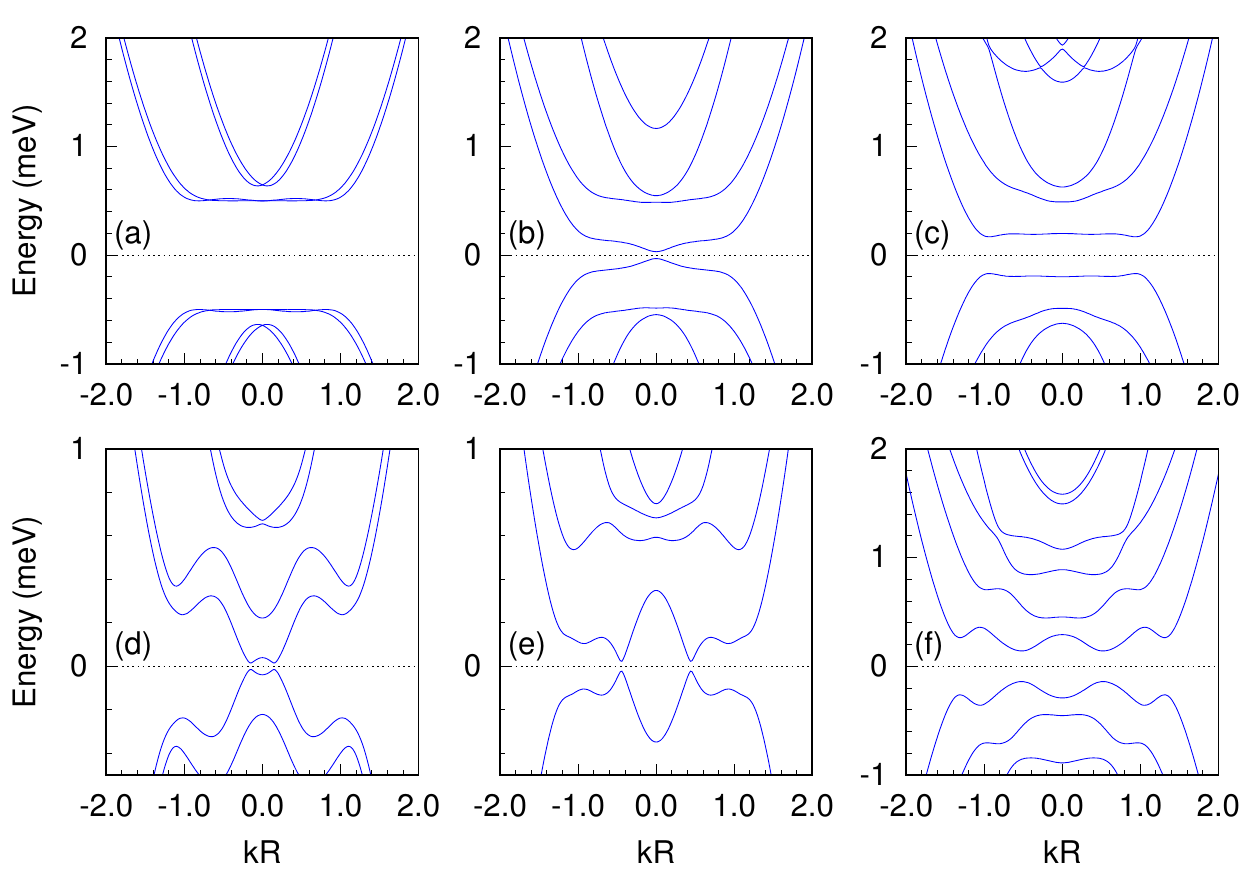}
\end{center}
\vspace{-5mm}
\caption{{\bf Hexagonal shell.} {BdG spectra, as in
Figs.~\ref{fig:EBdGk_tri}-\ref{fig:EBdGk_squ}, 
for a nanowire of radius $R=50$ nm and
shell thickness $t=9$ nm, i.\ e. ${\rm AR}=0.18$.}
(a) $\mu=286.9$ meV,\ $B=0$, 
(b) $\mu=286.9$ meV,\ $B=0.062$ T, 
(c) $\mu=286.9$ meV,\ $B=0.26$ T, 
(d) $\mu=287.3$ meV,\ $B=0.039$ T, 
(e) $\mu=287.3$ meV,\ $B=0.17$ T, 
(f) $\mu=287.3$ meV,\ $B=0.33$ T. 
}
\label{fig:EBdGk_hex}
\end{figure}

In Fig.\ \ref{fig:EBdGk_hex}, we show several spectra corresponding to
the hexagonal shell, now with six corner states involved.  The magnetic
fields necessary to close the superconductor gap has now lower values than
for the other geometries, a tendency that can also be observed for the
square vs. triangular case.  The reason is that the corner localization
softens when the angle of the polygons increase, and orbital effects
of the magnetic field are more important.  In other words, the split of
the energy bands occurs not only due to the spin Zeeman energy, but also
because of the orbital Zeeman energy.  We shall refer to these spectra
later when we shall discuss the phase diagrams. 

\begin{figure} [t]
\begin{center}
\includegraphics[width=85mm]{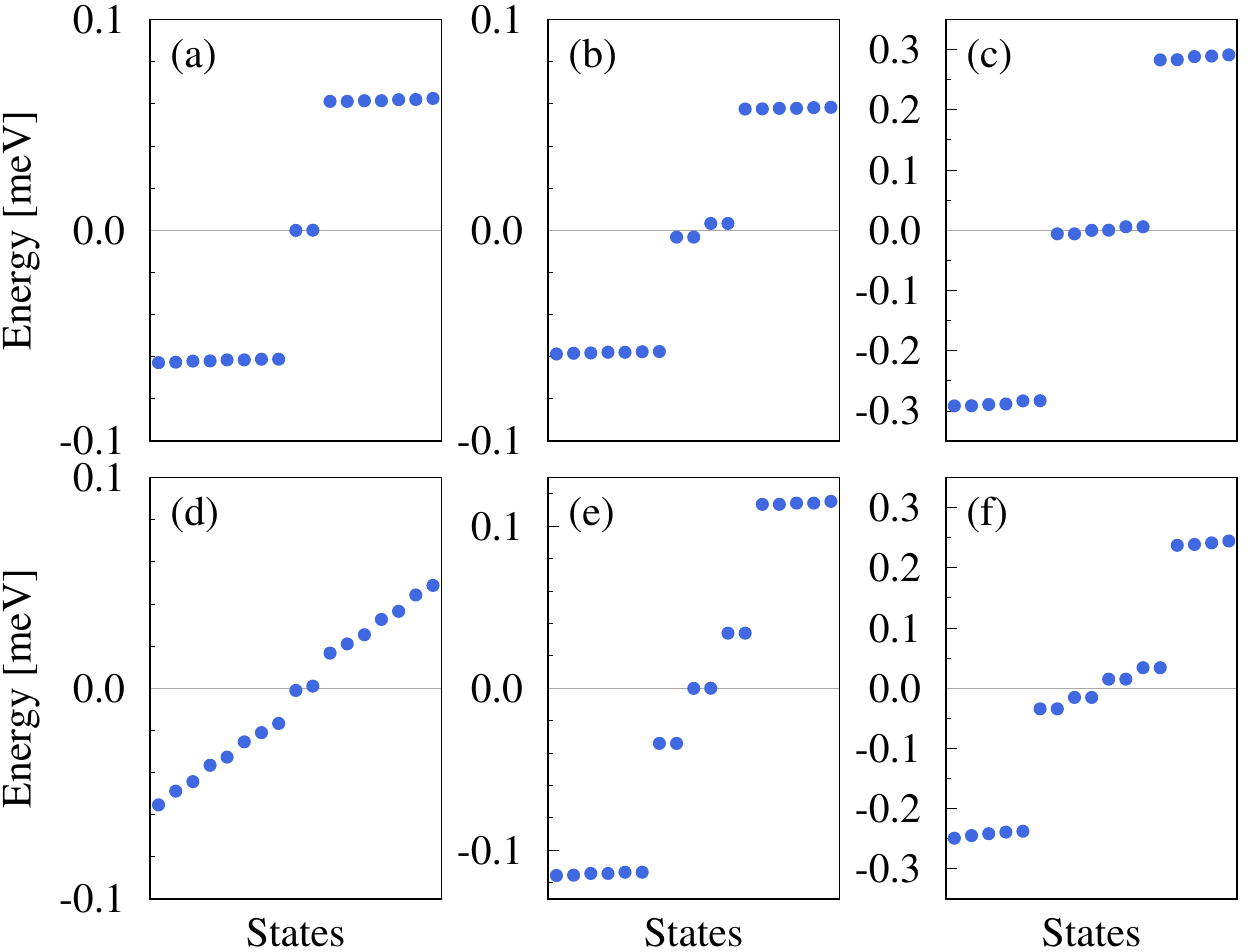}
\end{center}
\caption{Examples of 
BdG energy spectra for a nanowire of 10000 nm length and 50 nm radius
for selected values of the chemical potential and
Zeeman energy and two shell geometries: {\em triangular, the top row,} with 
thickness $t=9$ nm and
(a) $\mu=234.6$ meV, (b) $\mu=234.7$ meV, (c) $\mu=236.2$ meV, all with $B=1.32$ T;
{\em square, the bottom row,} with
(d)  $t=9$ nm, $\mu=277.6\ {\rm meV}\ B=0.33\ {\rm T}$, 
and also $t=8$ nm where
(e) $\mu=346.4\ {\rm meV}$ and
(f) $\mu=347.3\ {\rm meV}$ both with $B=0.66$~T. 
}
\label{fig:EBdG_fin}
\end{figure}

\subsection{BdG spectra for nanowires of finite length}

Next, we shall correlate the BdG spectra shown for infinite nanowires with 
some results for nanowires of finite length $L=200R=10000$ nm,
with triangular and square shells, shown in Fig.\ \ref{fig:EBdG_fin}.  
In the top row, Fig.\ \ref{fig:EBdG_fin}(a-c), we show three possible
situations obtained for the triangular geometry.  One pair of states at zero energy
is interpreted as a pair of Majorana (M) states, one at each end of the
wire, like in Fig.\ \ref{fig:EBdG_fin}(a).  For the infinite wire this
situation is shown in Fig.\ \ref{fig:EBdGk_tri}(b).  In this case the M
states are formed due to the particle-hole interaction within a single
corner state.  

When two corner states are involved at the same time, two particle-hole
symmetric pairs of states may be created, one pair with positive and
another one with negative energies, and both close to zero, as can be seen
in Fig.\ \ref{fig:EBdG_fin}(b). For the infinite nanowire this situation
corresponds to Fig.\ \ref{fig:EBdGk_tri}(c).  We shall call such states
pseudo (P) Majorana, and we shall denote the resulting combination of two
P pairs as PP.  We also notice that the gap between the other particle
and hole states is quite small in Figs. \ref{fig:EBdG_fin}(a-b), where
one or two corner states contribute to the low-energy states, respectively.

Fig.\ \ref{fig:EBdG_fin}(c) shows one M and two P states, and a
considerably larger gap.  We shall denote this combination as PMP.  In
this case all three corner states mix together.  The corresponding spectra
for the infinite wire have been shown in Fig.\ \ref{fig:EBdGk_tri}(d).
For a very thin shell, i.\ e.\ very low AR, the corner states would become
isolated from each other and each one would create an independent M state.
By slightly increasing the AR the wave functions of
different corner states develop a very small exponential overlap along the
sides of the polygonal shell, not visible in Fig.\ \ref{fig:polyloc}(a), 
which is equivalent to an interaction between the corner states.
This interaction lifts the degeneracy of the three former M states,
transforming two of them in P states, but leaving one M at zero energy,
hence giving a PMP configuration.  The energy of the P states can still be close
to zero if the energy separation between the corner states $\gamma$ is smaller that the
superconductivity parameter $\Delta$ such that the pairing interaction mixes
all corner states, provided the chemical potential is such that they
are all populated.  For the triangular shell with $t=9$ nm thickness we
have $\gamma_{\rm T}=0.027$ meV.

For the square geometry an M state exists in the situation shown in
Fig.\ \ref{fig:EBdGk_squ}(c) and a PP combination in that of Fig.\
\ref{fig:EBdGk_squ}(d).  The case of Fig.\ \ref{fig:EBdGk_squ}(b) with
vanishing gap at $kR\approx 0.3$ is shown in Fig.\ \ref{fig:EBdG_fin}(d).
A kind of M precursor state can be observed, eventually becoming a real
M state for a nanowire longer that 10000 nm.  The PMP configuration of
Fig.\ \ref{fig:EBdG_fin}(e) corresponds to Fig.\ \ref{fig:EBdGk_squ}(e).
Finally, in Fig.\ \ref{fig:EBdG_fin}(f) we obtain four pairs around zero
energy, to be called a 2PP combination, indicating four corner states
competing to create Majorana states. Again, like for the triangular case,
for a very low AR, the square shell yields four independent
or degenerate Majoranas at zero energy, whose degeneracy is lifted at
finite AR.  But unlike the triangular case, now, due to the particle-hole
symmetry, no pure Majorana can survive out of the former four, and
the result is a 2PP group, with energy dispersion smaller than the
superconductor gap.

\subsection{Semi-infinite nanowires}

\begin{figure} [t]
\begin{center}
\includegraphics[width=85mm]{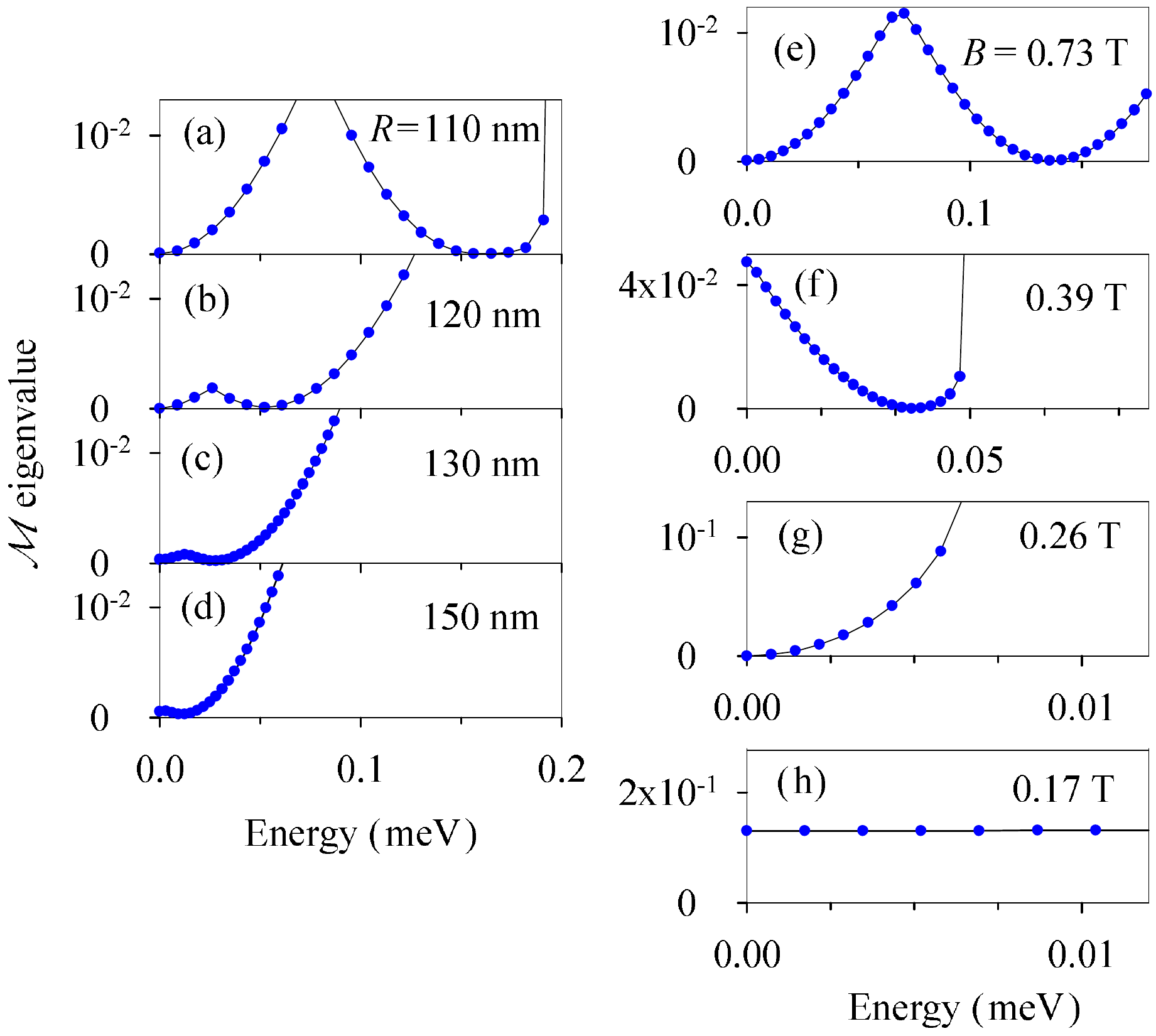}
\end{center}
\caption{Energy dependence of the lowest eigenvalue of matrix ${\cal M}$
in a semi-infinite triangular wire.
The zeroes indicate the physically acceptable energies for states attached to $z=0$ 
while decaying for increasing $z$.
Panels (a) to (d) correspond to a PMP configuration with increasing 
shell radius and fixed $t=40$ nm and $B=0.73$ T. 
Panels (e) to (h) correspond to a fixed $R=110$ nm and $t=37.5$ nm, and decreasing magnetic fields,
with configurations PMP (e), PP (f), M (g) and trivial (h).
Other parameters: $\mu = 22.7\, {\rm meV}$, 
 $\alpha=50\,{\rm meV}{\rm nm}$,
$\Delta=0.5\,{\rm meV}$. 
}
\label{fig:rads}
\end{figure}

We have confirmed the above scenario of zero-energy (M) and near-zero-energy (P)
pairs of states in the limit of very long wires 
directly studying the 
semi-infinite system. In this limit, any longitudinal finite size effect due to
interference between opposite wire ends is totally removed 
and we can unambiguously assign 
the fragmentation of the P pairs to the interaction between
the localized states along the edges of the prism. The semi-infinite 
system has been described with the complex-$k$ approach 
introduced in Ref.\ \onlinecite{ser13} for the case of 2D planar wires.
In this approach the existence of a state attached to the 
semi-infinite system boundary is signaled by a zero eigenvalue of a matrix ${\cal M}_{kk'}$,
defined by the set of evanescent modes $\{ k\}$ as labeled by their  
complex wavenumbers.

In Fig.\ \ref{fig:rads} we show the lowest eigenvalue of matrix 
${\cal M}$ as a function of the energy. Since the results are symmetric by
reverting the sign of $E$ we only show the positive energies.  Figures
\ref{fig:rads}(a-d) show the evolution with increasing shell radius $R$ of the
zeros of matrix ${\cal M}$ for a selected case. The chosen parameters
correspond to a PMP configuration in a triangular wire, with an ${\cal M}$
zero exactly at zero energy and two additional ones at finite energies
(a corresponding zero at negative energy is not shown).  By increasing
$R$ the prism edges become more and more independent, causing a collapse
towards zero energy of all three states.  Figures \ref{fig:rads}(e-h)
show the variation of the null eigenvalue for decreasing magnetic field
with a fixed radius.  The chosen shell width is rather large for a better
separation of the split P modes.  Increasing the field the configurations
in Fig.\ \ref{fig:rads} are: trivial (h), M (g), PP (f), PMP (e).

\begin{figure} [t]
\begin{center}
\includegraphics[width=75mm]{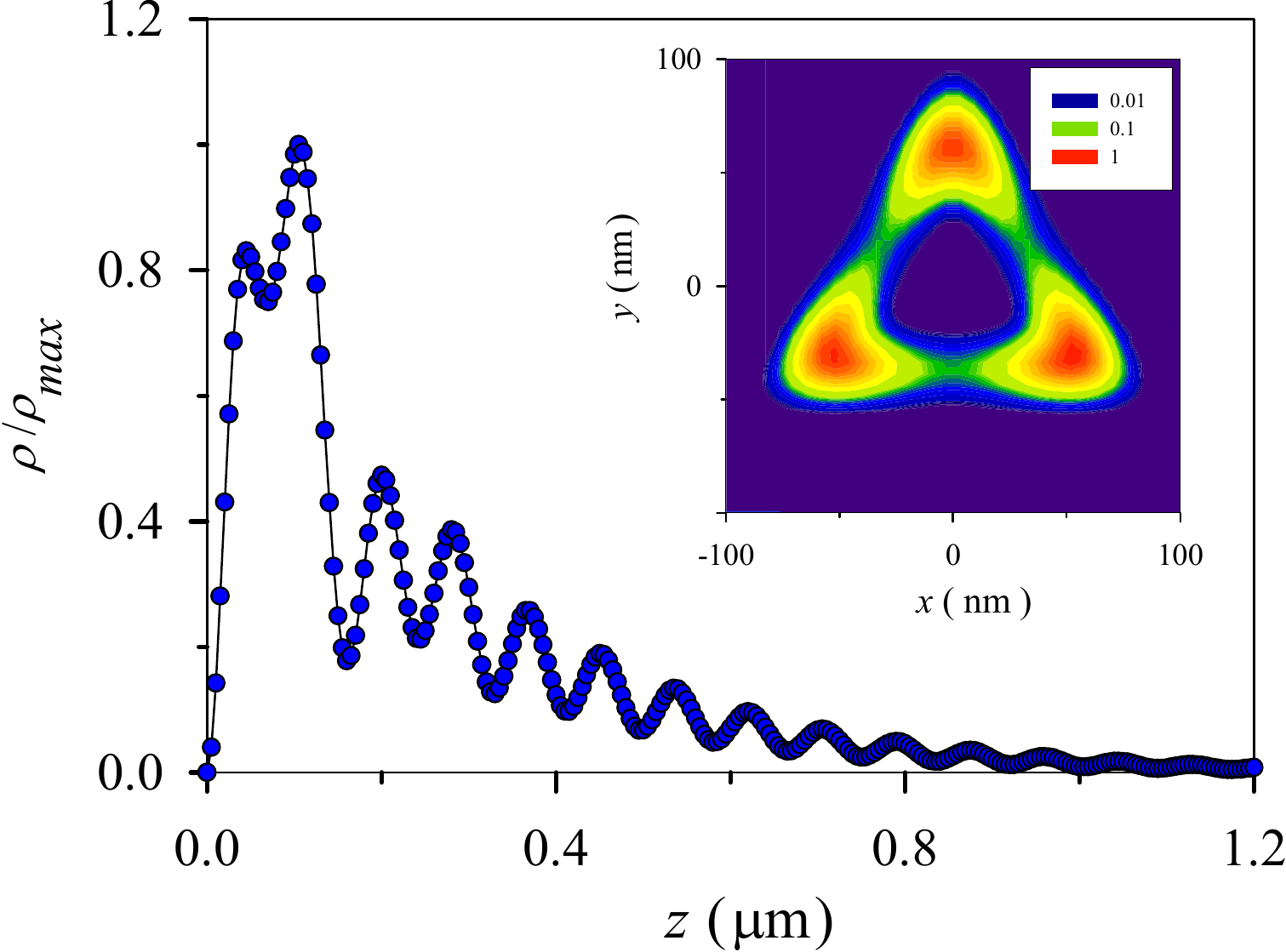}
\end{center}
\caption{Density corresponding to the M state of Fig.\ \ref{fig:rads}(a)
(at zero energy).  Dots show the $z$ evolution of the 1D density
integrated in the transverse directions, while the inset shows in a
color scale the transverse pattern for $z= 150\; {\rm nm}$, near the
maximum of the integrated 1D density.  }
\label{fig:dens}
\end{figure}

A characteristic density of an M mode is shown in Fig.\ \ref{fig:dens}.
The transverse-integrated density shows a shape similar to the known behavior from
purely 1D models,
with a decay length of around 1 $\mu$m for the chosen parameters. 
As anticipated, the transverse pattern exhibits localization on the edges of the triangular wire,
with sizable overlaps on the sides representative of edge-edge interactions in 
small enough wires ($R=110$ nm).

\section{\label{phase}Phase diagrams}

In this section we show phase diagrams in the parameter space $\mu$--$B$
obtained by calculating the minimum gap of the BdG energy spectra for the
infinite nanowire, at all $k$ values.  At any chemical potential $\mu$ the
phase transition from the trivial (no Majorana) to the topological state
(with Majorana) is expected to occur when the magnetic field is strong
enough to close the superconductor gap at $k=0$.  In principle the minimum
or critical magnetic field necessary for closing the superconductor gap
corresponds to a spin Zeeman energy $E_Z=2\Delta=1\ {\rm meV}$.  For our
material parameters this means $B=0.33$ T.  This result is however true
as long as the main effect of the magnetic field longitudinal to the
nanowire is to create only a spin splitting, and no orbital energy, e.\ g.\
for 1D or 2D (flat) nanowires.  For a prismatic core-shell nanowire this
can happen only if the corner states are almost isolated.  In our case
this situation occurs for the narrow triangular shell. 

\begin{figure} [t]
\includegraphics[width=85 mm]{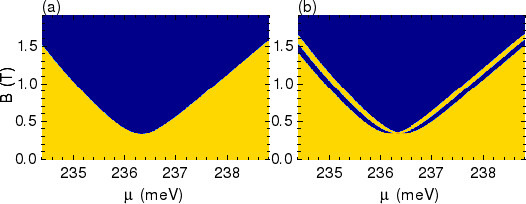} 

\vspace{1 mm}     
\includegraphics[width=85 mm]{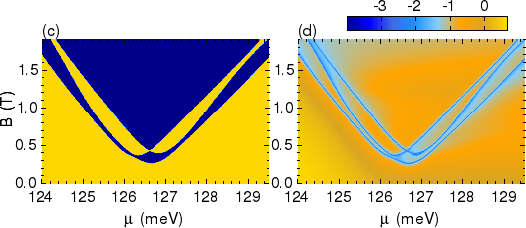}
\caption{{\bf Triangular shell} with $R=50$ nm.  In panels (a)-(c) the dark 
blue color indicates the topological Majorana phases and the yellow color
the trivial phases, the frontiers being defined by the gap closed at $k=0$.  
In (d) the colors indicate the minimum gap at all $k$ values, 
on the $\log_{10}$ scale. 
(a) Symmetric triangle with $t=9$ nm,  with corner states almost three-fold degenerated.  
(b) The same triangle in the presence of the weak transverse electric field which 
removes the degeneracy such that the phase boundary splits in three.
(c-d) A thicker shell, with $t=12.5$ nm. 
} 
\label{fig:Pdtri}
\end{figure}

In Fig.\ \ref{fig:Pdtri}(a) we show the phase diagram for the triangular
symmetric nanowire, where the topological phase is shown with
the dark blue color and the trivial phase with yellow.
By breaking the symmetry of the triangle, with the weak transversal
electric field, the phase boundary splits into three frontiers, each
one generated by different corner states, Fig.\ \ref{fig:Pdtri}(b).
The frontiers indicate the gap closed at $k=0$.
In both cases the topological phase above all lines is a PMP type
presented in Fig.\ \ref{fig:EBdG_fin}(c),  with low-energy P
states, implied by the small energy dispersion of the corner states 
$\gamma_{\rm T} < \Delta$.

In addition, for the asymmetric triangle, at any fixed chemical potential,
a succession of phases occurs when the magnetic field is increased
from zero.  After crossing the first frontier one enters a topological
state of an M type, illustrated in Fig.\ \ref{fig:EBdG_fin}(a). Then,
after crossing the second border a PP phase is obtained, [Fig.\
\ref{fig:EBdG_fin}(b)], which is in principle a topologically trivial
phase.  However these low-energy P states are expected to converge to
two Majorana pairs in the limit of a very small ratio $t/R$, 
and the phase to become topological.

Increasing now the thickness, to $t=12.5$ nm, the energy dispersion
of the corner states increases to $\gamma_{\rm T}=0.56$ meV and the
three frontiers become more separated, as seen in Fig.\ \ref{fig:Pdtri}(c).  
The energy interval between the inner and outer frontier is close to 
$\gamma_{\rm T}$.  Moreover, the orbital effects of the magnetic field
increase and the critical magnetic field slightly reduces to $B=0.27$ T.
But, the gaps can possibly shrink at nonzero $k$ values, as shown in
the example of Fig.\ \ref{fig:EBdGk_tri}(e), corresponding to the bottom of the 
topological phase, for $\mu=126.7$ meV and $B=0.34$ T.  Such small energy gaps
at $k\neq 0$ indicate the possible instability of the Majorana states.
Therefore, to incorporate that information, in Fig.\ \ref{fig:Pdtri}(d)
we repeat the phase diagram on a color scale indicating the minimum gap at
any $k$.  In this case the gaps are still reasonably large in most of the
regions such that the main topological phase is still robust.  One can
see for example the BdG spectrum corresponding to
$\mu=126.7$ meV and $B=1.32$ T shown in Fig.\ \ref{fig:EBdGk_tri}(f).
The spectrum for the finite wire is now qualitatively like in Fig.\
\ref{fig:EBdG_fin}(e), with a clear M state, but with the former P
states now with larger energy, that will further increase by increasing
the thickness of the shell.

\begin{figure} [t]
\includegraphics[width=85 mm]{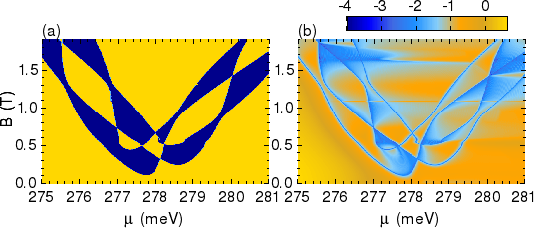}
\includegraphics[width=85 mm]{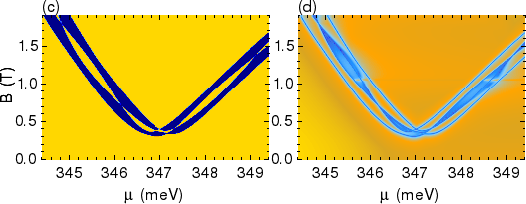}
\caption{{\bf Square shell} with $R=50$ nm. 
The topological (Majorana) and trivial phases are shown in dark blue 
and yellow, respectively, with frontiers defined by the gap closed 
at $k=0$, whereas the smallest gap at all $k$ values is shown with 
continuous colors (on $\log_{10}$ scale), like before.  
(a-b) $t=9$ nm, (c-d) $t=8$ nm.
}
\label{fig:Pdsqu}
\end{figure}

For the square shell with thickness $t=9$ nm the phase diagrams are
shown in Fig.\ \ref{fig:Pdsqu}.  We see now four phase boundaries,
corresponding to the four corner states, separated by an energy
$\gamma_{\rm S}$.  At the bottom of the phase diagram we notice a lower
critical magnetic field than for the triangular shell, of about 0.1 T, a
consequence of the increased orbital Zeeman energy.  But at the same time,
in the topological phase, the gaps at finite $k$ values can be small.
For example for the magnetic field $B=0.33$ T at $\mu=277.6$ meV the BdG
spectrum for the infinite wire is shown in Fig.\ \ref{fig:EBdGk_squ}(b)
and indicates a very small gap at $kR\approx 0.3$. Still, in the version
with finite length, Fig.\ \ref{fig:EBdG_fin}(d), we see two energies close
to zero, which would become an M state for a sufficiently long nanowire.
Indeed, that phase region is theoretically a topologically nontrivial one,
and in other regions the gap at finite $k$ may increase.  For example,
slightly to the left, for $\mu=276.5$ meV and $B=0.79$ T, the spectrum
is shown in Fig.\ \ref{fig:EBdGk_squ}(c) and it has a finite length
counterpart similar to that shown in Fig.\ \ref{fig:EBdG_fin}(a).

By crossing now each frontier defined by the gap closed at $k=0$, with
magnetic field increasing from zero, topologically nontrivial phases
containing M states alternate with topologically trivial phases
containing at most P states, depending on whether the number of crossed
frontiers is odd or even, respectively.  For example, with $\mu=276.5$
meV and $B=1.3$ T, the spectrum is that of Fig.\ \ref{fig:EBdGk_squ}(d),
with a PP configuration in the finite case.  Thus, the
phase surrounded by all four frontiers, situated in the middle of Fig.\
\ref{fig:Pdsqu}(a), is a trivial phase, where we expect only P states.

By reducing the aspect ratio of the polygon, for example by reducing the
thickness by only one nanometer the energy spread of the corner states rapidly
drops to $\gamma_{\rm S}=0.56$ meV (at $B=0$) and the phase boundaries 
approach each other, Fig.\ \ref{fig:Pdsqu}(c).   The central region contains
states of 2PP type, one of which being shown in Fig.\ \ref{fig:EBdG_fin}(f) for
$\mu=347.3$ meV and $B=0.66$ T.  Reducing the aspect ration of the square 
polygon further these states converge to four independent M
states.  Notice also that the critical field increases relatively to the 
previous case of $t=9$ nm because the orbital Zeeman effect drops. 

\begin{figure} [t]
\includegraphics[width=85 mm]{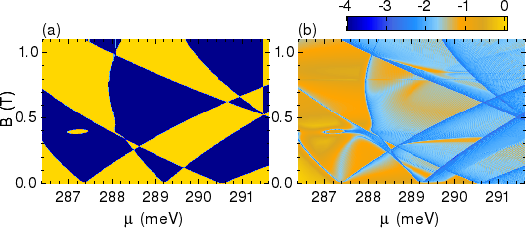}
\includegraphics[width=85 mm]{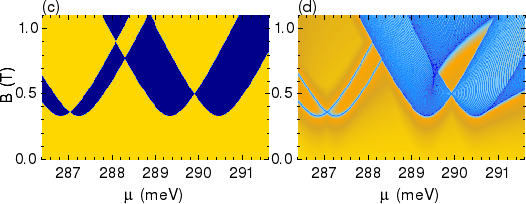}
\caption{{\bf Hexagonal shell} with $R=50$ nm and $t=9$ nm. 
(a-b) The phase diagrams are shown with the same color schemes as before
(Figs.\ \ref{fig:Pdtri}-\ref{fig:Pdsqu}).
(c-d) The results after excluding the orbital Zeeman effect of the magnetic field. 
}
\label{fig:Pdhex}
\end{figure} 

The orbital effect further increases for the hexagonal shell
with a similar aspect ratio as the other polygons, $t=9$ nm, such that
the critical magnetic field is about 12 mT for $\mu=287.3$ meV, Fig.\
\ref{fig:Pdhex}(a).  That phase boundary is created by the corner state
with the lowest energy.  As for the square geometry, the energy gap
in the topological phase may possibly be almost closed at nonzero $k$
as shown in Figs.\ \ref{fig:EBdGk_hex}(d) and (e), and also indicated
by the blue colors of Fig.\ \ref{fig:Pdhex}(b).  Still, robust
Majorana regions exists in this example at chemical potentials below 288 meV.
BdG spectra in one of such zones, at $\mu=286.8$ meV, are shown in Fig.\
\ref{fig:EBdGk_hex}(b) and (c), for $B=0.062$ T when the gap closes at
$k=0$ (i.\ e.\ near the phase boundary), and at $B=0.26$ T, when M states
are present. In the later case the eigenstates of a nanowire of finite
length are qualitatively similar to that of Fig.\ \ref{fig:EBdG_fin}(a).

In general both orbital and spin Zeeman energies may lead to
the splitting of electron and hole bands of the superconductor, which
would combine into M states at a sufficiently large magnetic field.
It turns out that for our hexagonal core-shell wire the orbital
splitting is dominant.  Consequently, in this case, the M states can
be obtained even if we neglect the $g$-factor.  With $g_{\rm eff}=0$
we could obtain a critical field of about $60$ mT.  This is an important
detail, since the $g$-factor in the shell material may be different,
possibly lower than that of the same bulk material.  A ``strange" detail of the phase
diagram can be the small elliptic island seen on the left side of Fig.\
\ref{fig:Pdhex}(b), around $\mu\approx 287.2$ meV and $B\approx
0.4$ T.  The reason for it is that, due to the combined orbital and spin
Zeeman splitting, the energy levels at $k=0$, forming the M states, can
possibly increase with the magnetic field, instead of decreasing, as expected
from the spin Zeeman effect alone.  Within that island the topological
phase is suppressed, although it exists all around outside it.

Finally, to close this section, 
in Fig.\ \ref{fig:Pdhex}(c-d) we show the phase diagram
for the hexagon after removing from the Hamiltonian (\ref{eq:Hw}) all 
terms related to the orbital Zeeman effect of the magnetic field,
to compare with the real phase diagram.  Indeed, the critical field
returns to the value expected from the spin Zeeman splitting only.
But with increasing the chemical potential the gap can easily close at 
finite $k$ values in the topological phase even without orbital effects.

\section{\label{conc}Summary and conclusions}


In conclusion, we have shown that prismatic core-shell nanowires
with proximity-induced superconductivity and Rashba spin-orbit coupling
provide an interesting and rather complex playground for Majorana
physics. A key new element that characterizes these structures is that the
Majorana physics can be realized within a low-energy sub-space defined
by states localized along the edges of the prism, which are separated
by a significant energy gap from higher energy states. Furthermore,
the localization of these states around the edges can be controlled,
e.\ g., by varying the thickness of the shell or of the core. Strong edge
localization leads to a system of effectively decoupled 1D nanowires
running along the edges of the prism, each wire hosting Majorana bound
states generated according to the well-known mechanism when the magnetic
field exceeds a certain critical value. Remarkably, upon increasing
the aspect ratio AR\! =\! $t/R$ the ``edge'' wires become coupled and an
interference between states localized on different edges emerges. In the
phase diagram, this edge-edge ``communication'' is manifested as a
fragmentation of the boundaries between trivial and topological
regions, as well as a separation of the Majorana bound states into
real Majoranas and ``trivial'', but nearly-zero energy modes
which we dub {\em pseudo Majorana} states.

The transformation of Majorana into pseudo Majorana states can be seen
as a finite size effect in the direction transversal to the nanowire.
The resulting phase diagrams correspond to  ladders of interacting
chains,\cite{Poyhonen14,Wakatsuki14,Sedlmayr16} as we demonstrate in the 
Appendix.

In the presence of  edge-edge interference there are regions in the $\mu{\rm
-}B$ phase diagram containing $n_p$ near-zero-energy pairs of states,
where $n_p$ can take all integer values from zero to the number of
prism edges. We have focussed on triangles and squares, exploring also
the results for hexagons.  Regions with odd values of $n_p$ contain a
genuine Majorana mode at zero energy, while those with even values of
$n_p$ only contain zero modes asymptotically, for low AR.
Actually, the energies of all the $n_p$ pairs collapse to zero energy in
the limit of low AR.  We stress that the Rashba interaction
in our model is not oriented along any externally fixed direction,
but points perpendicular to the sides and in a radial direction on the
edges. Therefore, we do not need to invoke a breaking of the overall
nanowire symmetry.  For given values of the shell thickness and radius, the
edge localization weakens when the number of corners is
increased. This is supported by our results for triangle, square and
hexagon. Indeed, the triangle has a relatively weaker fragmentation
of the phase diagram and of the pseudo Majorana pairs, while this
fragmentation is larger for the hexagon.

A relevant effect in core-shell nanowires is the possibility of
controlling the magneto-orbital contributions by varying the shell
thickness $t$.  Orbital effects of the magnetic field cause a strong
reduction of the global energy gap for propagating states at arbitrary
values of the wave vector $k$.  They can even lead to a complete closing of
the gap at a finite $k$ in some cases.  We have shown that in prismatic
core-shell wires with small values of $t$ orbital effects are greatly
quenched and, as a consequence, the energy gaps are sizeable larger than
for wide shell nanowires. This is a very appealing feature, since in
any practical application a sizeable gap is required for the stability
and protection of the Majorana states.

Our analysis uncovered a remarkable consequence of the magnetic orbital
effect in prismatic core-shell nanowires. Namely, the strong reduction of
the overall minimum magnetic field for a phase transition, for an arbitrary chemical
potential $\mu$.  This is a consequence of edge-edge interference in
thick enough core-shell nanowires.  For the hexagon this reduction is much
larger than for the triangular and square, due to the  enhanced edge-edge
interference. As mentioned, thick wires also tend to rapidly close the
overall gap for propagating states when increasing $B$. Still, our results
indicate that in hexagonal prismatic nanowires a compromise regime can
be found with a thickness $t$ such that the Majorana pairs are obtained,
along with a sizable overall gap, at low magnetic fields. Finally, as a general conclusion,
we emphasize that edge localization in prismatic core-shell nanowires
offers a promising physical mechanism towards the controllability of
Majorana states in nanowires.

\begin{acknowledgments}
This work was partially financed by the research funds of Reykjavik University.  
L.S. was supported by grant FIS2014-52564 (Mineco, Spain).
TDS was supported in part by NSF DMR-1414683.
\end{acknowledgments}

\appendix*

\renewcommand{\thefigure}{A\arabic{figure}}
\setcounter{figure}{0}
\setcounter{equation}{0}

\section{\label{toy}Toy model for proximitized core-shell nanowires}

\begin{figure}[t]
\begin{center}
\includegraphics[width=0.35\textwidth]{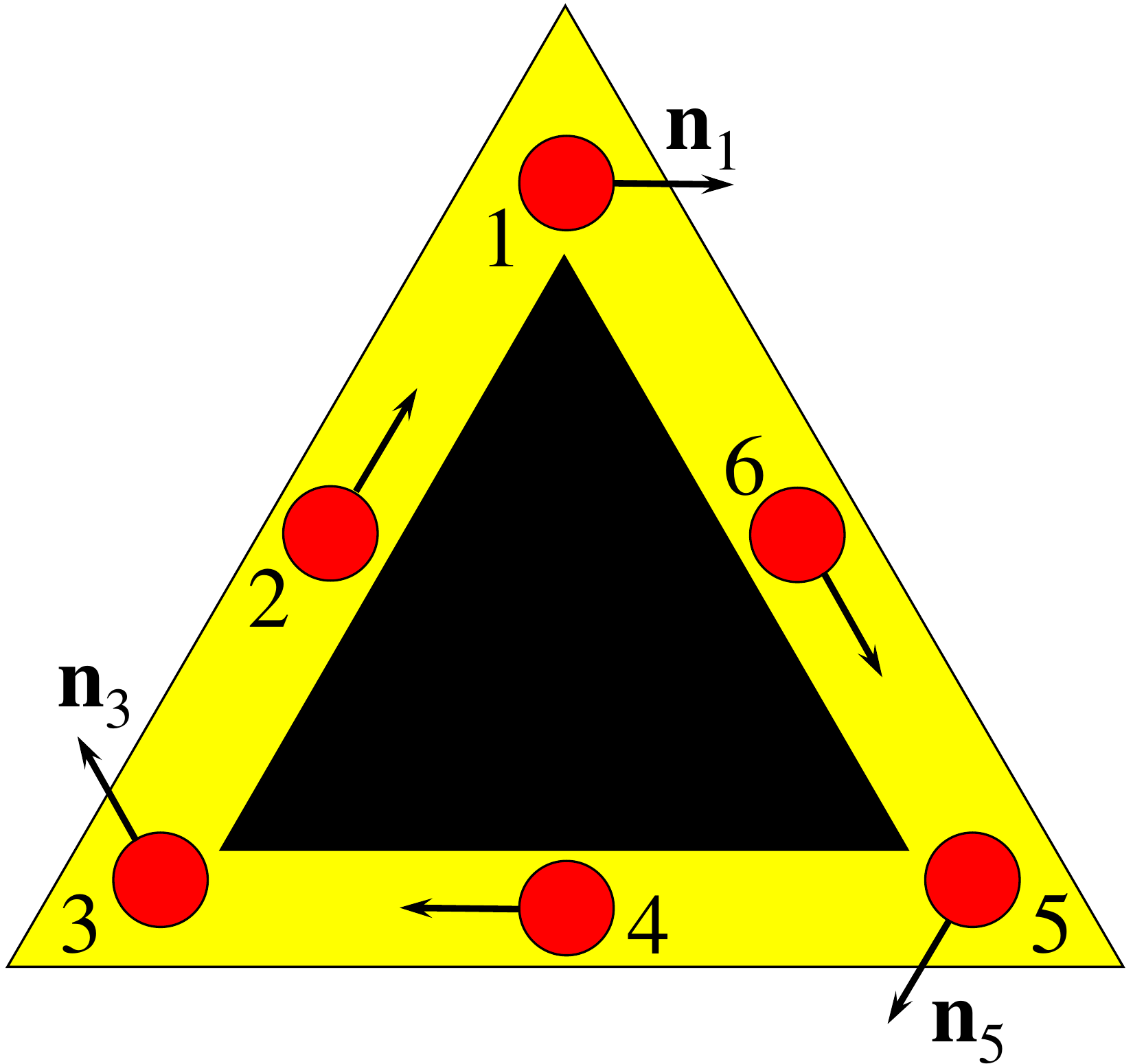}
\vspace{-3mm}
\end{center}
\caption{Schematic representation of the toy model
construction for a triangular wire. The shell (yellow/light gray) is
coarse-grained so that the vertices and the sides are represented by
one-dimensional chains (red/dark gray circles). The arrows indicate the
direction of the effective magnetic field ${\bm n}_\ell$ associated
with the (longitudinal) Rashba spin-orbit coupling.}
\vspace{-2mm}
\label{Fig0T}
\end{figure}

To gain further insight into the low-energy physics of proximitized
core-shell nanowires, and to better understand the qualitative dependence
of the BdG spectrum on relevant parameters, it is
convenient to use a simplified tight-binding ``toy model'' consisting of
coupled parallel chains. The basic idea is to define a ``coarse-grained''
shell consisting of one chain associated with each vertex and one chain
corresponding to each side, as illustrated in Fig.\ \ref{Fig0T}. If the
cross section of the wire is a polygon with $N$ corners, 
the toy model will contain $N$ or $2N$ parallel chains, depending on whether
only the corner or also the side states are included, respectively. 

Consider now a core-shell nanowire, described in a 
tight-binding manner in
the longitudinal direction, and proximity-coupled to one ore more
s-wave superconductors. 
The low-energy physics of the hybrid structure is described by the
following BdG Hamiltonian
\begin{eqnarray}
H&=&-t\!\sum_{i, \ell, \sigma} \!\left(c_{i+1 \ell \sigma}^\dagger  c_{i \ell \sigma}+h.c.\right)-t^\prime\!\sum_{i, \ell, \sigma} \!\left(c_{i \ell+1 \sigma}^\dagger  c_{i \ell \sigma}+h.c.\right) \nonumber \\
&+&\sum_{i, \ell, \sigma}\left[ V_{{\rm eff}}(\ell) - \mu\right]c_{i \ell \sigma}^\dagger  c_{i \ell \sigma}+\Gamma\sum_{i, \sigma}\sum_{\ell}^{(even)}c_{i \ell \sigma}^\dagger  c_{i \ell \sigma} \label{eq1T} \\
&+&\frac{i}{2}\sum_{i, \ell}\left[\alpha~\! c_{i+1 \ell}^\dagger \left(\hat{\bm{\sigma}}\cdot {\bm{n}}_\ell\right) c_{i \ell} +\alpha^\prime ~\! c_{i \ell+1}^\dagger  \hat{\sigma}_x c_{i \ell} +h.c.\right] \nonumber \\
&+& E_Z \sum_{i, \ell} c_{i \ell}^\dagger  \hat{\sigma}_x c_{i \ell} + \sum_{i, \ell} \left(\Delta_\ell c_{i \ell \uparrow}^\dagger c_{i \ell \downarrow}^\dagger + \Delta_\ell^*c_{i \ell \downarrow}  c_{i \ell \uparrow} \right), \nonumber
\end{eqnarray}
where $c_{i \ell \sigma}$ is the annihilation operator for an electron
with spin projection $\sigma$ localized on the longitudinal lattice site $i$ of
the chain $\ell$, $1\leq \ell \leq 2N$,  
and $c_{i \ell}=\left(c_{i \ell \uparrow}, 
c_{i \ell\downarrow}\right)^T$ 
is the corresponding spinor operator. The first term
in Eq.\ (\ref{eq1T}) represents nearest-neighbor hopping along the chains,
while the second term corresponds to the inter-chain coupling. 
In the summations over the chain index $\ell$ we use the convention
$2N+1\equiv 1$. The inter-chain coupling $t^\prime$ contains information
about the shell thickness, so that $t^\prime/t\rightarrow 0$ in the thin
shell limit.   The third term of the Hamiltonian (\ref{eq1T}) contains
the chemical potential $\mu$ and a chain-dependent effective potential
$V_{{\rm eff}}(\ell)$ that incorporates electrostatic effects due to back gates,
coupled superconductors, and non-homogeneous charge distributions. 
In general, $V_{{\rm eff}}(\ell)$ breaks the $N$-fold rotational symmetry
of the wire, and in addition it can also vary along the chains,
i.\ e., $V_{{\rm eff}} = V_{{\rm eff}}(i, \ell)$, but here we do not consider this
aspect.

The term proportional to $\Gamma$ accounts for the
property that the ``side states'' have higher energies than the ``corner
states'' as shown in Fig.\ \ref{fig:polyene}. The next term models a Rashba-type
SOI, with longitudinal and transverse components proportional to 
$\alpha$ and $\alpha^\prime$, respectively, generated by an effective 
electric field in the shell, Fig.\ \ref{fig:Rfield}. The corresponding  
direction of the effective magnetic field, ${\bm n}_\ell$, for
electrons moving along a triangular wire, is shown in Fig.\ \ref{Fig0T}. 
%
%
The last two terms in Eq.\ (\ref{eq1T}) describe the Zeeman splitting 
$E_Z= g_{\rm eff} \mu_B B$ generated by an external magnetic field 
applied along the chains, and the proximity-induced pairing. Note that
pairing potential $\Delta_\ell$ can be chain-dependent, which reflects the
generic situation when the surface of the wire is not uniformly covered by
a superconducting layer. Here we shall consider a constant pairing potential, 
$\Delta=0.3~$meV.
%
%
The other model parameters used in the numerical calculations presented
below have the following values: $t=5.64~$meV, $t^\prime=1.41~$meV,
$\alpha=2.0~$meV, $\alpha^\prime=0.5~$meV.
In these examples only the corner chains are considered, i.\ e. with 
odd $\ell$, corresponding to the previous calculations where the chemical 
potential was always inside the gap $\Gamma$.

\begin{figure}[t]
\begin{center}
\includegraphics[width=0.48\textwidth]{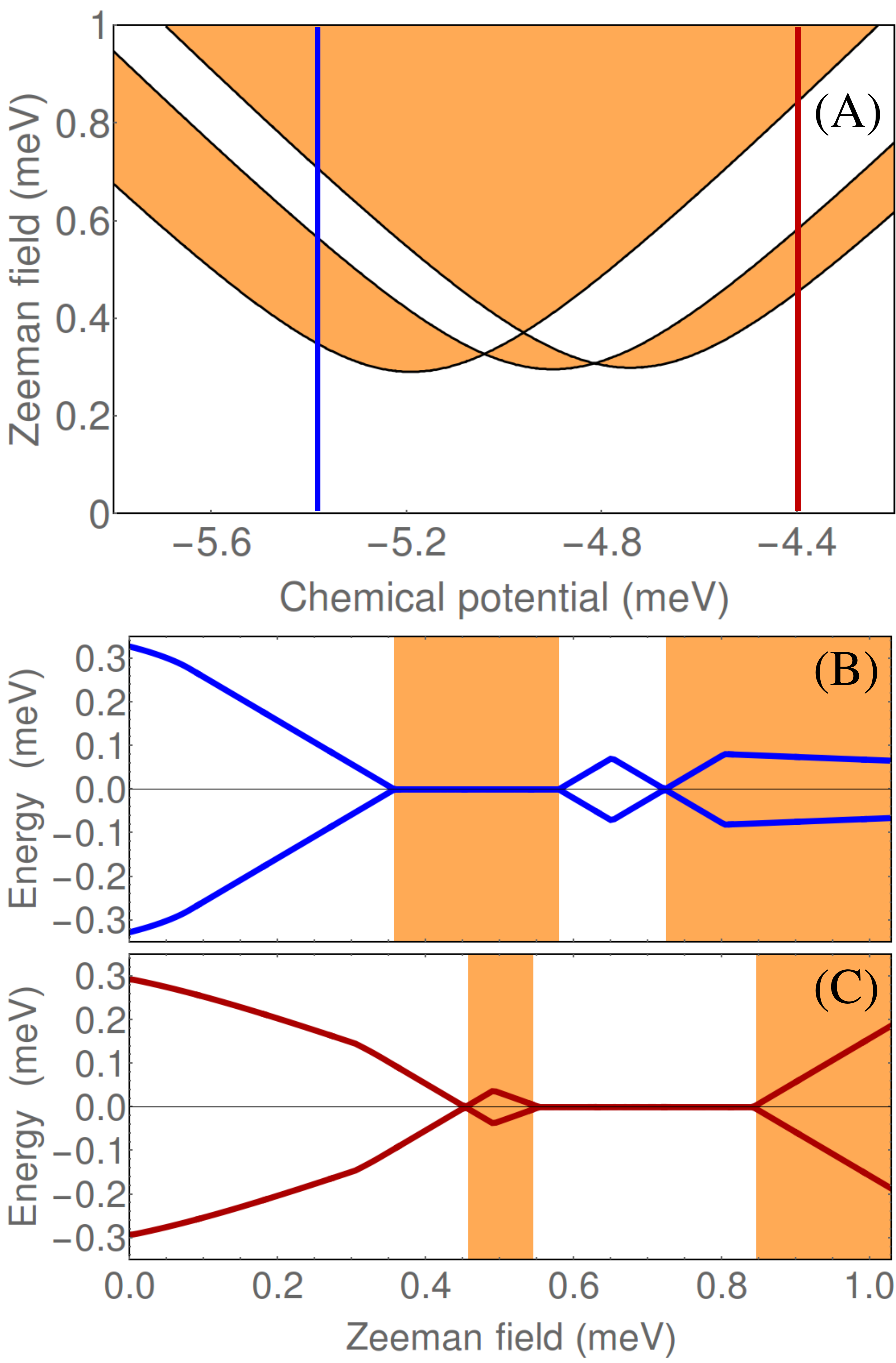}
\vspace{-3mm}
\end{center}
\caption{(A) Topological phase diagram for a symmetric triangular wire,
$V_{\rm eff}=0$. The white areas are topologically
trivial, while the orange (gray) regions correspond to ${\mathcal
M}=-1$. (B) Dependence of the minimum quasiparticle gap on the Zeeman
field for $\mu=-5.4~$meV [blue cut in panel (A)]. (C) Dependence of
the minimum quasiparticle gap on the Zeeman field for $\mu=-4.4~$meV
[dark red cut in panel (A)]. The white/orange regions correspond to the
phases shown in panel (A). Note the vanishing of the quasiparticle gap
in certain parameter regions.}
\vspace{-2mm}
\label{Fig1T}
\end{figure}

The first question that we address using the simplified tight-binding
model given by Eq.\ (\ref{eq1T}) concerns the structure of the phase
diagram of the proximitized wire. For this purpose we consider
a translation-invariant system (i.\ e., an infinite wire) and, for
concreteness, we focus on the triangular case, i.\ e.\ on the case $N=3$. 
To distinguish
between the topologically trivial and nontrivial phases, we calculate
the ${\mathbb Z}_2$ topological index ${\mathcal M}$ (the so-called {\em
Majorana number}) defined as\cite{Kitaev01}
\begin{equation}
{\mathcal M} = {\rm sign}\left[{\rm Pf}~\! B(0)\right]{\rm sign}\left[{\rm Pf}~\! B(\pi)\right],
\end{equation}
where ${\rm Pf}[\dots]$ designate the Pfaffian and $B(k)$ is a
momentum-dependent antisymmetric matrix (see below). The trivial phase
is characterized by ${\mathcal M} = +1$, while ${\mathcal M} = -1$
signals a topological superconducting phase. The antisymmetric matrix
$B$ represents the Fourier transform of the Hamiltonian (\ref{eq1T})
in the Majorana basis and  can be constructed\cite{Lutchyn10,Ghosh2010}
by virtue of the particle-hole symmetry of the BdG Hamiltonian,
\begin{equation}
{\mathcal T}{\mathcal H}(k){\mathcal T}^{-1} = {\mathcal H}(-k),
\end{equation}
where $H(k)$ is the Fourier transform of the (single particle) Hamiltonian
corresponding to Eq.\ (\ref{eq1T}) and ${\mathcal T} = U_t K$ is the
antiunitary time-reversal operator, with $U_t$  being a unitary operator
and $K$ the complex conjugation. One can easily verify that $B(k) =
{\mathcal H}(k) U_t$ is an antisymmetric matrix when calculated at
the time-reversal invariant points $k=0$ and $k=\pi$. Furthermore,
for physically-relevant model parameters (i.\ e., Zeeman splittings
not exceeding a few meVs and chemical potentials near the bottom of
the spectrum) we have ${\rm sign}\left[{\rm Pf}~\! B(\pi)\right]=+1$
and the  topological phase boundary is determined by a sign change of
${\rm Pf}~\! B(0)$. Note that ${\rm Det} {\mathcal H}(0) = \left[{\rm
Pf}~\! B(0)\right]^2$, hence the phase boundary is associated with the
vanishing of the quasiparticle gap at $k=0$.

The phase diagram for a triangular wire with $V_{{\rm eff}}(\ell)=0$ (i.\ e., no
symmetry-breaking potential) 
is shown in panel (A) of Fig.\ \ref{Fig1T}.
The white regions are characterized by ${\mathcal M}=+1$, i.\ e., they
are topologically trivial, while the orange (gray) areas correspond to
${\mathcal M}=-1$. 
The three phase boundaries would merge into a single one,
like in Fig.\ \ref{fig:Pdtri}(a), if $t^\prime/t\rightarrow 0$.
For a complementary characterization of different
phases, we calculate, as before, the minimum quasiparticle energy
(for all wave vectors), 
%
%
as shown in Fig.\ \ref{Fig1T}(B-C), 
for two different values of the chemical potential. 
At $E_Z=0$ the system is in a trivial superconducting phase
characterized by a quasiparticle gap close to $\Delta$. Increasing
$E_Z$ reduces the quasiparticle gap, which eventually vanishes
at a certain critical field. In panel (C), the gap reopens as we
enter a topological superconducting region (orange). By contrast,
in panel (B) the spectrum remains gapless throughout the first region
characterized by ${\mathcal M}=-1$, which means that the system  is a
gapless superconductor. Note that the vanishing of the gap happens at
$k\neq 0$, except for the phase boundary crossing points, where the
quasiparticle gap vanishes at $k=0$, as mentioned above.  A gapless
superconducting phase is also present in panel (C) for Zeeman fields
between approximately $0.55~$meV and $0.85~$meV, where ${\mathcal
M}=+1$. Finally, for large-enough values of the Zeeman splitting,
i.\ e. above $0.7~$meV in panel (B) and $0.85~$meV in panel (C), the system
is in a  gapped topological phase.

\begin{figure}[t]
\begin{center}
\includegraphics[width=0.48\textwidth]{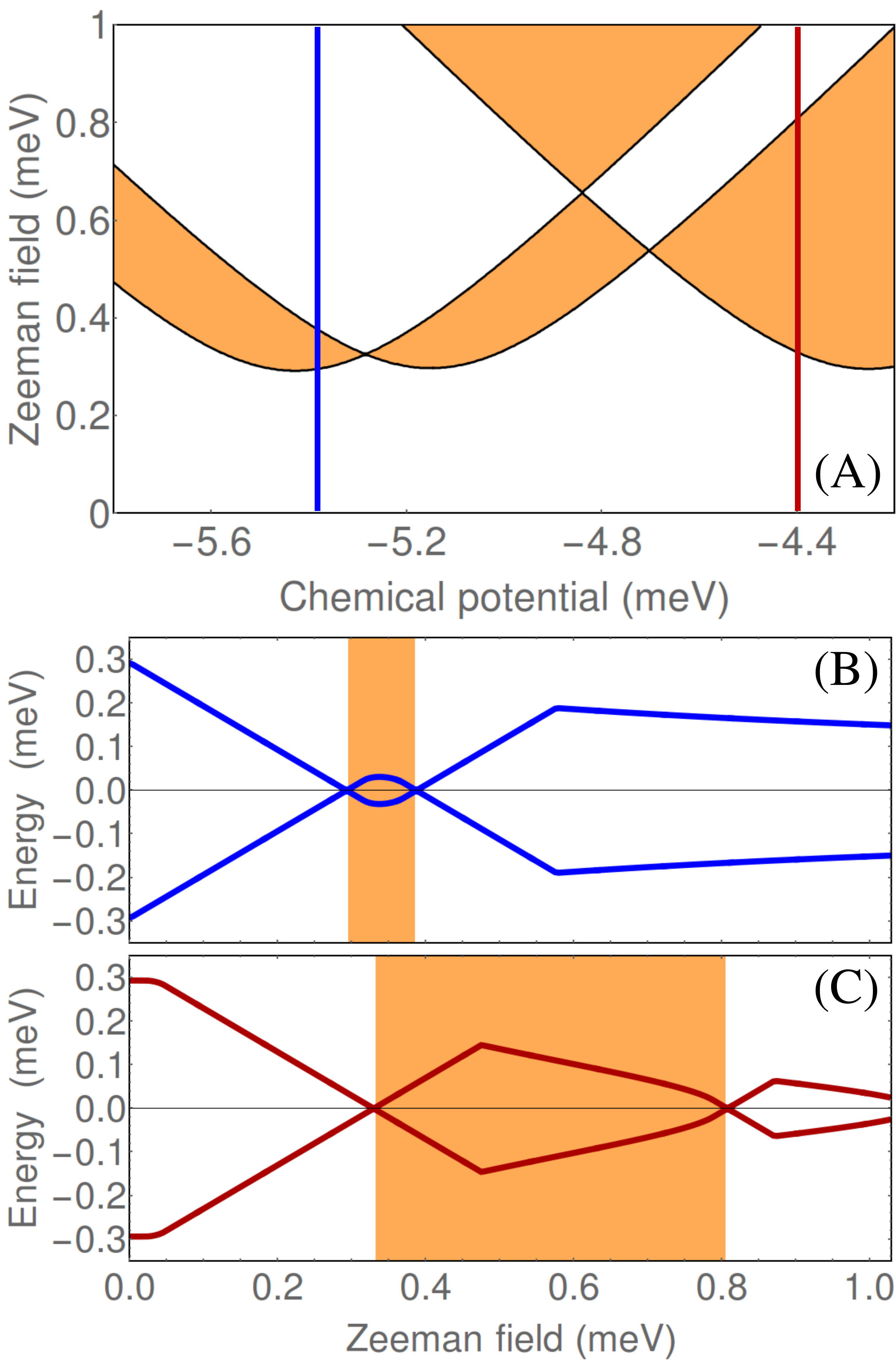}
\vspace{-3mm}
\end{center}
\caption{(A) Topological phase diagram as function of the chemical potential and applied Zeeman field for a slightly asymmetric triangular wire, with
$V_{{\rm eff}}= (0.67, -0.33, -0.33)$ meV at corners. The white and orange (gray) phases are topologically trivial and nontrivial, respectively. (B) Dependence of the minimum quasiparticle gap on the Zeeman field for $\mu=-5.4~$meV [blue cut in panel (A)]. (C) Dependence of the minimum quasiparticle gap on the Zeeman field for $\mu=-4.4~$meV [dark red cut in panel (A)].  Note that all superconducting phases are gapped.}
\vspace{-2mm}
\label{Fig2T}
\end{figure}

Next, we consider the effect of a symmetry-breaking potential 
$V_{{\rm eff}}(\ell) = (0.67, -0.33, -0.33)$ meV for $\ell=1,3,5$.
The results are shown in Fig.\ \ref{Fig2T}. First, we note that
the location of the phase boundaries changes significantly, but the
topology of the phase diagram remains the same. Second, upon breaking the
three-fold rotation symmetry of the wire the gapless superconducting
phases shown in Fig.\ \ref{Fig1T} become gapped. Finally, we note
that the low-field topological phase corresponding to $\mu=-4.4~$meV
is characterized by a sizable quasiparticle gap [see panel (C)],
which suggests that this regime may be particularly suitable for the
experimental realization of a topological superconducting state that
supports robust zero-energy  Majorana modes.


\begin{thebibliography}{50}%
\makeatletter
\providecommand \@ifxundefined [1]{%
 \@ifx{#1\undefined}
}%
\providecommand \@ifnum [1]{%
 \ifnum #1\expandafter \@firstoftwo
 \else \expandafter \@secondoftwo
 \fi
}%
\providecommand \@ifx [1]{%
 \ifx #1\expandafter \@firstoftwo
 \else \expandafter \@secondoftwo
 \fi
}%
\providecommand \natexlab [1]{#1}%
\providecommand \enquote  [1]{``#1''}%
\providecommand \bibnamefont  [1]{#1}%
\providecommand \bibfnamefont [1]{#1}%
\providecommand \citenamefont [1]{#1}%
\providecommand \href@noop [0]{\@secondoftwo}%
\providecommand \href [0]{\begingroup \@sanitize@url \@href}%
\providecommand \@href[1]{\@@startlink{#1}\@@href}%
\providecommand \@@href[1]{\endgroup#1\@@endlink}%
\providecommand \@sanitize@url [0]{\catcode `\\12\catcode `\$12\catcode
  `\&12\catcode `\#12\catcode `\^12\catcode `\_12\catcode `\%12\relax}%
\providecommand \@@startlink[1]{}%
\providecommand \@@endlink[0]{}%
\providecommand \url  [0]{\begingroup\@sanitize@url \@url }%
\providecommand \@url [1]{\endgroup\@href {#1}{\urlprefix }}%
\providecommand \urlprefix  [0]{URL }%
\providecommand \Eprint [0]{\href }%
\providecommand \doibase [0]{http://dx.doi.org/}%
\providecommand \selectlanguage [0]{\@gobble}%
\providecommand \bibinfo  [0]{\@secondoftwo}%
\providecommand \bibfield  [0]{\@secondoftwo}%
\providecommand \translation [1]{[#1]}%
\providecommand \BibitemOpen [0]{}%
\providecommand \bibitemStop [0]{}%
\providecommand \bibitemNoStop [0]{.\EOS\space}%
\providecommand \EOS [0]{\spacefactor3000\relax}%
\providecommand \BibitemShut  [1]{\csname bibitem#1\endcsname}%
\let\auto@bib@innerbib\@empty
\bibitem [{\citenamefont {Kitaev}(2001)}]{Kitaev01}%
  \BibitemOpen
  \bibfield  {author} {\bibinfo {author} {\bibfnamefont {A.~Y.}\ \bibnamefont
  {Kitaev}},\ }\href@noop {} {\bibfield  {journal} {\bibinfo  {journal}
  {Physics-Uspekhi}\ }\textbf {\bibinfo {volume} {44}},\ \bibinfo {pages} {131}
  (\bibinfo {year} {2001})}\BibitemShut {NoStop}%
\bibitem [{\citenamefont {Kitaev}(2003)}]{Kitaev03}%
  \BibitemOpen
  \bibfield  {author} {\bibinfo {author} {\bibfnamefont {A.~Y.}\ \bibnamefont
  {Kitaev}},\ }\href@noop {} {\bibfield  {journal} {\bibinfo  {journal} {Ann.
  Phys.}\ }\textbf {\bibinfo {volume} {303}},\ \bibinfo {pages} {2} (\bibinfo
  {year} {2003})}\BibitemShut {NoStop}%
\bibitem [{\citenamefont {Wilczek}(2009)}]{Wilczek09}%
  \BibitemOpen
  \bibfield  {author} {\bibinfo {author} {\bibfnamefont {F.}~\bibnamefont
  {Wilczek}},\ }\href@noop {} {\bibfield  {journal} {\bibinfo  {journal}
  {Nature Phys.}\ }\textbf {\bibinfo {volume} {5}},\ \bibinfo {pages} {614}
  (\bibinfo {year} {2009})}\BibitemShut {NoStop}%
\bibitem [{\citenamefont {Nayak}\ \emph {et~al.}(2008)\citenamefont {Nayak},
  \citenamefont {Simon}, \citenamefont {Stern}, \citenamefont {Freedman},\ and\
  \citenamefont {Das~Sarma}}]{Nayak08}%
  \BibitemOpen
  \bibfield  {author} {\bibinfo {author} {\bibfnamefont {C.}~\bibnamefont
  {Nayak}}, \bibinfo {author} {\bibfnamefont {S.~H.}\ \bibnamefont {Simon}},
  \bibinfo {author} {\bibfnamefont {A.}~\bibnamefont {Stern}}, \bibinfo
  {author} {\bibfnamefont {M.}~\bibnamefont {Freedman}}, \ and\ \bibinfo
  {author} {\bibfnamefont {S.}~\bibnamefont {Das~Sarma}},\ }\href@noop {}
  {\bibfield  {journal} {\bibinfo  {journal} {Rev. Mod. Phys.}\ }\textbf
  {\bibinfo {volume} {80}},\ \bibinfo {pages} {1083} (\bibinfo {year}
  {2008})}\BibitemShut {NoStop}%
\bibitem [{\citenamefont {Das~Sarma}\ \emph {et~al.}(2015)\citenamefont
  {Das~Sarma}, \citenamefont {Freedman},\ and\ \citenamefont
  {Nayak}}]{DSarma2015}%
  \BibitemOpen
  \bibfield  {author} {\bibinfo {author} {\bibfnamefont {S.}~\bibnamefont
  {Das~Sarma}}, \bibinfo {author} {\bibfnamefont {M.}~\bibnamefont {Freedman}},
  \ and\ \bibinfo {author} {\bibfnamefont {C.}~\bibnamefont {Nayak}},\
  }\href@noop {} {\bibfield  {journal} {\bibinfo  {journal} {Npj Quantum
  Information}\ }\textbf {\bibinfo {volume} {1}},\ \bibinfo {pages} {15001}
  (\bibinfo {year} {2015})}\BibitemShut {NoStop}%
\bibitem [{\citenamefont {Stanescu}(2017)}]{Stanescu2017}%
  \BibitemOpen
  \bibfield  {author} {\bibinfo {author} {\bibfnamefont {T.~D.}\ \bibnamefont
  {Stanescu}},\ }\href@noop {} {\emph {\bibinfo {title} {Introduction to
  topological quantum matter and quantum computation}}}\ (\bibinfo  {publisher}
  {CRC Press, Taylor \& Francis Group},\ \bibinfo {year} {2017})\BibitemShut
  {NoStop}%
\bibitem [{\citenamefont {Fu}\ and\ \citenamefont {Kane}(2008)}]{Fu08}%
  \BibitemOpen
  \bibfield  {author} {\bibinfo {author} {\bibfnamefont {L.}~\bibnamefont
  {Fu}}\ and\ \bibinfo {author} {\bibfnamefont {C.~L.}\ \bibnamefont {Kane}},\
  }\href@noop {} {\bibfield  {journal} {\bibinfo  {journal} {Phys. Rev. Lett.}\
  }\textbf {\bibinfo {volume} {100}},\ \bibinfo {pages} {096407} (\bibinfo
  {year} {2008})}\BibitemShut {NoStop}%
\bibitem [{\citenamefont {Sau}\ \emph {et~al.}(2010)\citenamefont {Sau},
  \citenamefont {Lutchyn}, \citenamefont {Tewari},\ and\ \citenamefont
  {Das~Sarma}}]{Sau10}%
  \BibitemOpen
  \bibfield  {author} {\bibinfo {author} {\bibfnamefont {J.~D.}\ \bibnamefont
  {Sau}}, \bibinfo {author} {\bibfnamefont {R.~M.}\ \bibnamefont {Lutchyn}},
  \bibinfo {author} {\bibfnamefont {S.}~\bibnamefont {Tewari}}, \ and\ \bibinfo
  {author} {\bibfnamefont {S.}~\bibnamefont {Das~Sarma}},\ }\href@noop {}
  {\bibfield  {journal} {\bibinfo  {journal} {Phys. Rev. Lett.}\ }\textbf
  {\bibinfo {volume} {104}},\ \bibinfo {pages} {040502} (\bibinfo {year}
  {2010})}\BibitemShut {NoStop}%
\bibitem [{\citenamefont {Nadj-Perge}\ \emph {et~al.}(2014)\citenamefont
  {Nadj-Perge}, \citenamefont {Drozdov}, \citenamefont {Li}, \citenamefont
  {Chen}, \citenamefont {Jeon}, \citenamefont {Seo}, \citenamefont {MacDonald},
  \citenamefont {Bernevig},\ and\ \citenamefont {Yazdani}}]{NadjPerge2014}%
  \BibitemOpen
  \bibfield  {author} {\bibinfo {author} {\bibfnamefont {S.}~\bibnamefont
  {Nadj-Perge}}, \bibinfo {author} {\bibfnamefont {I.~K.}\ \bibnamefont
  {Drozdov}}, \bibinfo {author} {\bibfnamefont {J.}~\bibnamefont {Li}},
  \bibinfo {author} {\bibfnamefont {H.}~\bibnamefont {Chen}}, \bibinfo {author}
  {\bibfnamefont {S.}~\bibnamefont {Jeon}}, \bibinfo {author} {\bibfnamefont
  {J.}~\bibnamefont {Seo}}, \bibinfo {author} {\bibfnamefont {A.~H.}\
  \bibnamefont {MacDonald}}, \bibinfo {author} {\bibfnamefont {B.~A.}\
  \bibnamefont {Bernevig}}, \ and\ \bibinfo {author} {\bibfnamefont
  {A.}~\bibnamefont {Yazdani}},\ }\href@noop {} {\bibfield  {journal} {\bibinfo
   {journal} {Science}\ }\textbf {\bibinfo {volume} {346}},\ \bibinfo {pages}
  {602} (\bibinfo {year} {2014})}\BibitemShut {NoStop}%
\bibitem [{\citenamefont {Alicea}(2012)}]{Alicea12}%
  \BibitemOpen
  \bibfield  {author} {\bibinfo {author} {\bibfnamefont {J.}~\bibnamefont
  {Alicea}},\ }\href@noop {} {\bibfield  {journal} {\bibinfo  {journal} {Rep.\
  Prog.\ Phys.}\ }\textbf {\bibinfo {volume} {75}},\ \bibinfo {pages} {076501}
  (\bibinfo {year} {2012})}\BibitemShut {NoStop}%
\bibitem [{\citenamefont {Stanescu}\ and\ \citenamefont
  {Tewari}(2013)}]{Stanescu13a}%
  \BibitemOpen
  \bibfield  {author} {\bibinfo {author} {\bibfnamefont {T.~D.}\ \bibnamefont
  {Stanescu}}\ and\ \bibinfo {author} {\bibfnamefont {S.}~\bibnamefont
  {Tewari}},\ }\href@noop {} {\bibfield  {journal} {\bibinfo  {journal} {J.
  Phys.: Condens. Matter}\ }\textbf {\bibinfo {volume} {25}},\ \bibinfo {pages}
  {233201} (\bibinfo {year} {2013})}\BibitemShut {NoStop}%
\bibitem [{\citenamefont {Beenakker}(2013)}]{Beenakker13}%
  \BibitemOpen
  \bibfield  {author} {\bibinfo {author} {\bibfnamefont {C.~W.~J.}\
  \bibnamefont {Beenakker}},\ }\href@noop {} {\bibfield  {journal} {\bibinfo
  {journal} {Annual Review of Cond. Matt. Phys.}\ }\textbf {\bibinfo {volume}
  {4}},\ \bibinfo {pages} {113} (\bibinfo {year} {2013})}\BibitemShut {NoStop}%
\bibitem [{\citenamefont {Franz}(2013)}]{Franz13}%
  \BibitemOpen
  \bibfield  {author} {\bibinfo {author} {\bibfnamefont {M.}~\bibnamefont
  {Franz}},\ }\href@noop {} {\bibfield  {journal} {\bibinfo  {journal} {Nature
  Nanotechnology}\ }\textbf {\bibinfo {volume} {8}},\ \bibinfo {pages} {149}
  (\bibinfo {year} {2013})}\BibitemShut {NoStop}%
\bibitem [{\citenamefont {Lutchyn}\ \emph {et~al.}(2010)\citenamefont
  {Lutchyn}, \citenamefont {Sau},\ and\ \citenamefont {Das~Sarma}}]{Lutchyn10}%
  \BibitemOpen
  \bibfield  {author} {\bibinfo {author} {\bibfnamefont {R.~M.}\ \bibnamefont
  {Lutchyn}}, \bibinfo {author} {\bibfnamefont {J.~D.}\ \bibnamefont {Sau}}, \
  and\ \bibinfo {author} {\bibfnamefont {S.}~\bibnamefont {Das~Sarma}},\
  }\href@noop {} {\bibfield  {journal} {\bibinfo  {journal} {Phys. Rev. Lett.}\
  }\textbf {\bibinfo {volume} {105}},\ \bibinfo {pages} {077001} (\bibinfo
  {year} {2010})}\BibitemShut {NoStop}%
\bibitem [{\citenamefont {Oreg}\ \emph {et~al.}(2010)\citenamefont {Oreg},
  \citenamefont {Refael},\ and\ \citenamefont {von Oppen}}]{Oreg10}%
  \BibitemOpen
  \bibfield  {author} {\bibinfo {author} {\bibfnamefont {Y.}~\bibnamefont
  {Oreg}}, \bibinfo {author} {\bibfnamefont {G.}~\bibnamefont {Refael}}, \ and\
  \bibinfo {author} {\bibfnamefont {F.}~\bibnamefont {von Oppen}},\ }\href@noop
  {} {\bibfield  {journal} {\bibinfo  {journal} {Phys. Rev. Lett.}\ }\textbf
  {\bibinfo {volume} {105}},\ \bibinfo {pages} {177002} (\bibinfo {year}
  {2010})}\BibitemShut {NoStop}%
\bibitem [{\citenamefont {Mourik}\ \emph {et~al.}(2012)\citenamefont {Mourik},
  \citenamefont {Zuo}, \citenamefont {Frolov}, \citenamefont {Plissard},
  \citenamefont {Bakkers},\ and\ \citenamefont {Kouwenhoven}}]{Mourik12}%
  \BibitemOpen
  \bibfield  {author} {\bibinfo {author} {\bibfnamefont {V.}~\bibnamefont
  {Mourik}}, \bibinfo {author} {\bibfnamefont {K.}~\bibnamefont {Zuo}},
  \bibinfo {author} {\bibfnamefont {S.}~\bibnamefont {Frolov}}, \bibinfo
  {author} {\bibfnamefont {S.}~\bibnamefont {Plissard}}, \bibinfo {author}
  {\bibfnamefont {E.}~\bibnamefont {Bakkers}}, \ and\ \bibinfo {author}
  {\bibfnamefont {L.}~\bibnamefont {Kouwenhoven}},\ }\href@noop {} {\bibfield
  {journal} {\bibinfo  {journal} {Science}\ }\textbf {\bibinfo {volume}
  {336}},\ \bibinfo {pages} {1003} (\bibinfo {year} {2012})}\BibitemShut
  {NoStop}%
\bibitem [{\citenamefont {Deng}\ \emph {et~al.}(2012)\citenamefont {Deng},
  \citenamefont {Yu}, \citenamefont {Huang}, \citenamefont {Larsson},
  \citenamefont {Caroff},\ and\ \citenamefont {Xu}}]{Deng12}%
  \BibitemOpen
  \bibfield  {author} {\bibinfo {author} {\bibfnamefont {M.}~\bibnamefont
  {Deng}}, \bibinfo {author} {\bibfnamefont {C.}~\bibnamefont {Yu}}, \bibinfo
  {author} {\bibfnamefont {G.}~\bibnamefont {Huang}}, \bibinfo {author}
  {\bibfnamefont {M.}~\bibnamefont {Larsson}}, \bibinfo {author} {\bibfnamefont
  {P.}~\bibnamefont {Caroff}}, \ and\ \bibinfo {author} {\bibfnamefont
  {H.}~\bibnamefont {Xu}},\ }\href@noop {} {\bibfield  {journal} {\bibinfo
  {journal} {Nano Lett.}\ }\textbf {\bibinfo {volume} {12}},\ \bibinfo {pages}
  {6414} (\bibinfo {year} {2012})}\BibitemShut {NoStop}%
\bibitem [{\citenamefont {Das}\ \emph {et~al.}(2012)\citenamefont {Das},
  \citenamefont {Ronen}, \citenamefont {Most}, \citenamefont {Oreg},
  \citenamefont {Heiblum},\ and\ \citenamefont {Shtrikman}}]{Das12}%
  \BibitemOpen
  \bibfield  {author} {\bibinfo {author} {\bibfnamefont {A.}~\bibnamefont
  {Das}}, \bibinfo {author} {\bibfnamefont {Y.}~\bibnamefont {Ronen}}, \bibinfo
  {author} {\bibfnamefont {Y.}~\bibnamefont {Most}}, \bibinfo {author}
  {\bibfnamefont {Y.}~\bibnamefont {Oreg}}, \bibinfo {author} {\bibfnamefont
  {M.}~\bibnamefont {Heiblum}}, \ and\ \bibinfo {author} {\bibfnamefont
  {H.}~\bibnamefont {Shtrikman}},\ }\href@noop {} {\bibfield  {journal}
  {\bibinfo  {journal} {Nature Phys.}\ }\textbf {\bibinfo {volume} {8}},\
  \bibinfo {pages} {887} (\bibinfo {year} {2012})}\BibitemShut {NoStop}%
\bibitem [{\citenamefont {Churchill}\ \emph {et~al.}(2013)\citenamefont
  {Churchill}, \citenamefont {Fatemi}, \citenamefont {Grove-Rasmussen},
  \citenamefont {Deng}, \citenamefont {Caroff}, \citenamefont {Xu},\ and\
  \citenamefont {Marcus}}]{Churchill13}%
  \BibitemOpen
  \bibfield  {author} {\bibinfo {author} {\bibfnamefont {H.}~\bibnamefont
  {Churchill}}, \bibinfo {author} {\bibfnamefont {V.}~\bibnamefont {Fatemi}},
  \bibinfo {author} {\bibfnamefont {K.}~\bibnamefont {Grove-Rasmussen}},
  \bibinfo {author} {\bibfnamefont {M.}~\bibnamefont {Deng}}, \bibinfo {author}
  {\bibfnamefont {P.}~\bibnamefont {Caroff}}, \bibinfo {author} {\bibfnamefont
  {H.}~\bibnamefont {Xu}}, \ and\ \bibinfo {author} {\bibfnamefont
  {C.}~\bibnamefont {Marcus}},\ }\href@noop {} {\bibfield  {journal} {\bibinfo
  {journal} {Phys. Rev. B}\ }\textbf {\bibinfo {volume} {87}},\ \bibinfo
  {pages} {241401} (\bibinfo {year} {2013})}\BibitemShut {NoStop}%
\bibitem [{\citenamefont {Finck}\ \emph {et~al.}(2013)\citenamefont {Finck},
  \citenamefont {Van~Harlingen}, \citenamefont {Mohseni}, \citenamefont
  {Jung},\ and\ \citenamefont {Li}}]{Finck13}%
  \BibitemOpen
  \bibfield  {author} {\bibinfo {author} {\bibfnamefont {A.~D.~K.}\
  \bibnamefont {Finck}}, \bibinfo {author} {\bibfnamefont {D.~J.}\ \bibnamefont
  {Van~Harlingen}}, \bibinfo {author} {\bibfnamefont {P.~K.}\ \bibnamefont
  {Mohseni}}, \bibinfo {author} {\bibfnamefont {K.}~\bibnamefont {Jung}}, \
  and\ \bibinfo {author} {\bibfnamefont {X.}~\bibnamefont {Li}},\ }\href@noop
  {} {\bibfield  {journal} {\bibinfo  {journal} {Phys. Rev. Lett.}\ }\textbf
  {\bibinfo {volume} {110}},\ \bibinfo {pages} {126406} (\bibinfo {year}
  {2013})}\BibitemShut {NoStop}%
\bibitem [{\citenamefont {Chang}\ \emph {et~al.}(2015)\citenamefont {Chang},
  \citenamefont {Albrecht}, \citenamefont {Jespersen}, \citenamefont
  {Kuemmeth}, \citenamefont {Krogstrup}, \citenamefont {Nyg{\r a}rd},\ and\
  \citenamefont {Marcus}}]{Chang2015}%
  \BibitemOpen
  \bibfield  {author} {\bibinfo {author} {\bibfnamefont {W.}~\bibnamefont
  {Chang}}, \bibinfo {author} {\bibfnamefont {S.~M.}\ \bibnamefont {Albrecht}},
  \bibinfo {author} {\bibfnamefont {T.~S.}\ \bibnamefont {Jespersen}}, \bibinfo
  {author} {\bibfnamefont {F.}~\bibnamefont {Kuemmeth}}, \bibinfo {author}
  {\bibfnamefont {P.}~\bibnamefont {Krogstrup}}, \bibinfo {author}
  {\bibfnamefont {J.}~\bibnamefont {Nyg{\r a}rd}}, \ and\ \bibinfo {author}
  {\bibfnamefont {C.~M.}\ \bibnamefont {Marcus}},\ }\href@noop {} {\bibfield
  {journal} {\bibinfo  {journal} {Nat Nano}\ }\textbf {\bibinfo {volume}
  {10}},\ \bibinfo {pages} {232} (\bibinfo {year} {2015})}\BibitemShut
  {NoStop}%
\bibitem [{\citenamefont {Krogstrup}\ \emph {et~al.}(2015)\citenamefont
  {Krogstrup}, \citenamefont {Ziino}, \citenamefont {Chang}, \citenamefont
  {Albrecht}, \citenamefont {Madsen}, \citenamefont {Johnson}, \citenamefont
  {Nyg{\r a}rd}, \citenamefont {Marcus},\ and\ \citenamefont
  {Jespersen}}]{Krogstrup2015}%
  \BibitemOpen
  \bibfield  {author} {\bibinfo {author} {\bibfnamefont {P.}~\bibnamefont
  {Krogstrup}}, \bibinfo {author} {\bibfnamefont {N.~L.~B.}\ \bibnamefont
  {Ziino}}, \bibinfo {author} {\bibfnamefont {W.}~\bibnamefont {Chang}},
  \bibinfo {author} {\bibfnamefont {S.~M.}\ \bibnamefont {Albrecht}}, \bibinfo
  {author} {\bibfnamefont {M.~H.}\ \bibnamefont {Madsen}}, \bibinfo {author}
  {\bibfnamefont {E.}~\bibnamefont {Johnson}}, \bibinfo {author} {\bibfnamefont
  {J.}~\bibnamefont {Nyg{\r a}rd}}, \bibinfo {author} {\bibfnamefont {C.~M.}\
  \bibnamefont {Marcus}}, \ and\ \bibinfo {author} {\bibfnamefont {T.~S.}\
  \bibnamefont {Jespersen}},\ }\href {\doibase 10.1038/nmat4176} {\bibfield
  {journal} {\bibinfo  {journal} {Nature Materials}\ }\textbf {\bibinfo
  {volume} {14}},\ \bibinfo {pages} {400} (\bibinfo {year} {2015})}\BibitemShut
  {NoStop}%
\bibitem [{\citenamefont {Albrecht}\ \emph {et~al.}(2016)\citenamefont
  {Albrecht}, \citenamefont {Higginbotham}, \citenamefont {Madsen},
  \citenamefont {Kuemmeth}, \citenamefont {Jespersen}, \citenamefont {Nyg{\r
  a}rd}, \citenamefont {Krogstrup},\ and\ \citenamefont
  {Marcus}}]{Albrecht2016}%
  \BibitemOpen
  \bibfield  {author} {\bibinfo {author} {\bibfnamefont {S.~M.}\ \bibnamefont
  {Albrecht}}, \bibinfo {author} {\bibfnamefont {A.~P.}\ \bibnamefont
  {Higginbotham}}, \bibinfo {author} {\bibfnamefont {M.}~\bibnamefont
  {Madsen}}, \bibinfo {author} {\bibfnamefont {F.}~\bibnamefont {Kuemmeth}},
  \bibinfo {author} {\bibfnamefont {T.~S.}\ \bibnamefont {Jespersen}}, \bibinfo
  {author} {\bibfnamefont {J.}~\bibnamefont {Nyg{\r a}rd}}, \bibinfo {author}
  {\bibfnamefont {P.}~\bibnamefont {Krogstrup}}, \ and\ \bibinfo {author}
  {\bibfnamefont {C.~M.}\ \bibnamefont {Marcus}},\ }\href@noop {} {\bibfield
  {journal} {\bibinfo  {journal} {Nature}\ }\textbf {\bibinfo {volume} {531}},\
  \bibinfo {pages} {206} (\bibinfo {year} {2016})}\BibitemShut {NoStop}%
\bibitem [{\citenamefont {Deng}\ \emph {et~al.}(2016)\citenamefont {Deng},
  \citenamefont {Vaitiekenas}, \citenamefont {Hansen}, \citenamefont {Danon},
  \citenamefont {Leijnse}, \citenamefont {Flensberg}, \citenamefont {Nyg{\r
  a}rd}, \citenamefont {Krogstrup},\ and\ \citenamefont {Marcus}}]{Deng2016}%
  \BibitemOpen
  \bibfield  {author} {\bibinfo {author} {\bibfnamefont {M.~T.}\ \bibnamefont
  {Deng}}, \bibinfo {author} {\bibfnamefont {S.}~\bibnamefont {Vaitiekenas}},
  \bibinfo {author} {\bibfnamefont {E.~B.}\ \bibnamefont {Hansen}}, \bibinfo
  {author} {\bibfnamefont {J.}~\bibnamefont {Danon}}, \bibinfo {author}
  {\bibfnamefont {M.}~\bibnamefont {Leijnse}}, \bibinfo {author} {\bibfnamefont
  {K.}~\bibnamefont {Flensberg}}, \bibinfo {author} {\bibfnamefont
  {J.}~\bibnamefont {Nyg{\r a}rd}}, \bibinfo {author} {\bibfnamefont
  {P.}~\bibnamefont {Krogstrup}}, \ and\ \bibinfo {author} {\bibfnamefont
  {C.~M.}\ \bibnamefont {Marcus}},\ }\href@noop {} {\bibfield  {journal}
  {\bibinfo  {journal} {Science}\ }\textbf {\bibinfo {volume} {354}},\ \bibinfo
  {pages} {1557} (\bibinfo {year} {2016})}\BibitemShut {NoStop}%
\bibitem [{\citenamefont {Zhang}\ \emph {et~al.}(2016)\citenamefont {Zhang},
  \citenamefont {G\"ul}, \citenamefont {Conesa-Boj}, \citenamefont {Zuo},
  \citenamefont {Mourik}, \citenamefont {de~Vries}, \citenamefont {van Veen},
  \citenamefont {van Woerkom}, \citenamefont {Nowak}, \citenamefont {Wimmer},
  \citenamefont {Car}, \citenamefont {Plissard}, \citenamefont {Bakkers},
  \citenamefont {Quintero-P\'erez}, \citenamefont {Goswami}, \citenamefont
  {Watanabe}, \citenamefont {Taniguchi},\ and\ \citenamefont
  {Kouwenhoven}}]{Zhang16}%
  \BibitemOpen
  \bibfield  {author} {\bibinfo {author} {\bibfnamefont {H.}~\bibnamefont
  {Zhang}}, \bibinfo {author} {\bibfnamefont {O.}~\bibnamefont {G\"ul}},
  \bibinfo {author} {\bibfnamefont {S.}~\bibnamefont {Conesa-Boj}}, \bibinfo
  {author} {\bibfnamefont {K.}~\bibnamefont {Zuo}}, \bibinfo {author}
  {\bibfnamefont {V.}~\bibnamefont {Mourik}}, \bibinfo {author} {\bibfnamefont
  {F.~K.}\ \bibnamefont {de~Vries}}, \bibinfo {author} {\bibfnamefont
  {J.}~\bibnamefont {van Veen}}, \bibinfo {author} {\bibfnamefont {D.~J.}\
  \bibnamefont {van Woerkom}}, \bibinfo {author} {\bibfnamefont {M.~P.}\
  \bibnamefont {Nowak}}, \bibinfo {author} {\bibfnamefont {M.}~\bibnamefont
  {Wimmer}}, \bibinfo {author} {\bibfnamefont {D.}~\bibnamefont {Car}},
  \bibinfo {author} {\bibfnamefont {S.}~\bibnamefont {Plissard}}, \bibinfo
  {author} {\bibfnamefont {E.~P. A.~M.}\ \bibnamefont {Bakkers}}, \bibinfo
  {author} {\bibfnamefont {M.}~\bibnamefont {Quintero-P\'erez}}, \bibinfo
  {author} {\bibfnamefont {S.}~\bibnamefont {Goswami}}, \bibinfo {author}
  {\bibfnamefont {K.}~\bibnamefont {Watanabe}}, \bibinfo {author}
  {\bibfnamefont {T.}~\bibnamefont {Taniguchi}}, \ and\ \bibinfo {author}
  {\bibfnamefont {L.~P.}\ \bibnamefont {Kouwenhoven}},\ }\href@noop {}
  {\bibfield  {journal} {\bibinfo  {journal} {arXiv:1603.04069}\ } (\bibinfo
  {year} {2016})}\BibitemShut {NoStop}%
\bibitem [{\citenamefont {Chen}\ \emph {et~al.}(2016)\citenamefont {Chen},
  \citenamefont {Yu}, \citenamefont {Stenger}, \citenamefont {Hocevar},
  \citenamefont {Car}, \citenamefont {Plissard}, \citenamefont {Bakkers},
  \citenamefont {Stanescu},\ and\ \citenamefont {Frolov}}]{Chen2016}%
  \BibitemOpen
  \bibfield  {author} {\bibinfo {author} {\bibfnamefont {J.}~\bibnamefont
  {Chen}}, \bibinfo {author} {\bibfnamefont {P.}~\bibnamefont {Yu}}, \bibinfo
  {author} {\bibfnamefont {J.}~\bibnamefont {Stenger}}, \bibinfo {author}
  {\bibfnamefont {M.}~\bibnamefont {Hocevar}}, \bibinfo {author} {\bibfnamefont
  {D.}~\bibnamefont {Car}}, \bibinfo {author} {\bibfnamefont {S.~R.}\
  \bibnamefont {Plissard}}, \bibinfo {author} {\bibfnamefont {E.~P.}\
  \bibnamefont {Bakkers}}, \bibinfo {author} {\bibfnamefont {T.~D.}\
  \bibnamefont {Stanescu}}, \ and\ \bibinfo {author} {\bibfnamefont {S.~M.}\
  \bibnamefont {Frolov}},\ }\href@noop {} {\bibfield  {journal} {\bibinfo
  {journal} {e-print arXiv:1610.04555}\ } (\bibinfo {year} {2016})}\BibitemShut
  {NoStop}%
\bibitem [{\citenamefont {Stanescu}\ \emph {et~al.}(2011)\citenamefont
  {Stanescu}, \citenamefont {Lutchyn},\ and\ \citenamefont
  {Das~Sarma}}]{Stanescu11}%
  \BibitemOpen
  \bibfield  {author} {\bibinfo {author} {\bibfnamefont {T.~D.}\ \bibnamefont
  {Stanescu}}, \bibinfo {author} {\bibfnamefont {R.~M.}\ \bibnamefont
  {Lutchyn}}, \ and\ \bibinfo {author} {\bibfnamefont {S.}~\bibnamefont
  {Das~Sarma}},\ }\href@noop {} {\bibfield  {journal} {\bibinfo  {journal}
  {Phys. Rev. B}\ }\textbf {\bibinfo {volume} {84}},\ \bibinfo {pages} {144522}
  (\bibinfo {year} {2011})}\BibitemShut {NoStop}%
\bibitem [{\citenamefont {Nijholt}\ and\ \citenamefont
  {Akhmerov}(2016)}]{Nijholt16}%
  \BibitemOpen
  \bibfield  {author} {\bibinfo {author} {\bibfnamefont {B.}~\bibnamefont
  {Nijholt}}\ and\ \bibinfo {author} {\bibfnamefont {A.~R.}\ \bibnamefont
  {Akhmerov}},\ }\href@noop {} {\bibfield  {journal} {\bibinfo  {journal}
  {Phys. Rev. B}\ }\textbf {\bibinfo {volume} {93}},\ \bibinfo {pages} {235434}
  (\bibinfo {year} {2016})}\BibitemShut {NoStop}%
\bibitem [{\citenamefont {Ihn}(2010)}]{Ihn10}%
  \BibitemOpen
  \bibfield  {author} {\bibinfo {author} {\bibfnamefont {T.}~\bibnamefont
  {Ihn}},\ }\href@noop {} {\emph {\bibinfo {title} {Semiconductor
  nanostructures}}}\ (\bibinfo  {publisher} {Oxford},\ \bibinfo {year}
  {2010})\BibitemShut {NoStop}%
\bibitem [{\citenamefont {Maier}\ \emph {et~al.}(2014)\citenamefont {Maier},
  \citenamefont {Klinovaja},\ and\ \citenamefont {Loss}}]{Maier14}%
  \BibitemOpen
  \bibfield  {author} {\bibinfo {author} {\bibfnamefont {F.}~\bibnamefont
  {Maier}}, \bibinfo {author} {\bibfnamefont {J.}~\bibnamefont {Klinovaja}}, \
  and\ \bibinfo {author} {\bibfnamefont {D.}~\bibnamefont {Loss}},\ }\href
  {\doibase 10.1103/PhysRevB.90.195421} {\bibfield  {journal} {\bibinfo
  {journal} {Phys. Rev. B}\ }\textbf {\bibinfo {volume} {90}},\ \bibinfo
  {pages} {195421} (\bibinfo {year} {2014})}\BibitemShut {NoStop}%
\bibitem [{\citenamefont {Bertoni}\ \emph {et~al.}(2011)\citenamefont
  {Bertoni}, \citenamefont {Royo}, \citenamefont {Mahawish},\ and\
  \citenamefont {Goldoni}}]{Bertoni11}%
  \BibitemOpen
  \bibfield  {author} {\bibinfo {author} {\bibfnamefont {A.}~\bibnamefont
  {Bertoni}}, \bibinfo {author} {\bibfnamefont {M.}~\bibnamefont {Royo}},
  \bibinfo {author} {\bibfnamefont {F.}~\bibnamefont {Mahawish}}, \ and\
  \bibinfo {author} {\bibfnamefont {G.}~\bibnamefont {Goldoni}},\ }\href
  {\doibase 10.1103/PhysRevB.84.205323} {\bibfield  {journal} {\bibinfo
  {journal} {Phys. Rev. B}\ }\textbf {\bibinfo {volume} {84}},\ \bibinfo
  {pages} {205323} (\bibinfo {year} {2011})}\BibitemShut {NoStop}%
\bibitem [{\citenamefont {Wu}\ \emph {et~al.}(1992)\citenamefont {Wu},
  \citenamefont {Sprung},\ and\ \citenamefont {Martorell}}]{Wu92}%
  \BibitemOpen
  \bibfield  {author} {\bibinfo {author} {\bibfnamefont {H.}~\bibnamefont
  {Wu}}, \bibinfo {author} {\bibfnamefont {D.~W.~L.}\ \bibnamefont {Sprung}}, \
  and\ \bibinfo {author} {\bibfnamefont {J.}~\bibnamefont {Martorell}},\
  }\href@noop {} {\bibfield  {journal} {\bibinfo  {journal} {J. Appl. Phys.}\
  }\textbf {\bibinfo {volume} {72}},\ \bibinfo {pages} {151} (\bibinfo {year}
  {1992})}\BibitemShut {NoStop}%
\bibitem [{\citenamefont {Londergan}\ \emph {et~al.}(1999)\citenamefont
  {Londergan}, \citenamefont {Carini},\ and\ \citenamefont
  {Murdock}}]{Londergan99}%
  \BibitemOpen
  \bibfield  {author} {\bibinfo {author} {\bibfnamefont {J.~T.}\ \bibnamefont
  {Londergan}}, \bibinfo {author} {\bibfnamefont {J.~P.}\ \bibnamefont
  {Carini}}, \ and\ \bibinfo {author} {\bibfnamefont {D.~P.}\ \bibnamefont
  {Murdock}},\ }\href@noop {} {\emph {\bibinfo {title} {Binding and Scattering
  in Two-Dimensional Systems. Applications to Quantum Wires, Waveguides and
  Photonic Crystals}}}\ (\bibinfo  {publisher} {Springer},\ \bibinfo {year}
  {1999})\BibitemShut {NoStop}%
\bibitem [{\citenamefont {Hu}\ \emph {et~al.}(2011)\citenamefont {Hu},
  \citenamefont {Zhang}, \citenamefont {Giapis},\ and\ \citenamefont
  {Poulikakos}}]{Hu11}%
  \BibitemOpen
  \bibfield  {author} {\bibinfo {author} {\bibfnamefont {M.}~\bibnamefont
  {Hu}}, \bibinfo {author} {\bibfnamefont {X.}~\bibnamefont {Zhang}}, \bibinfo
  {author} {\bibfnamefont {K.~P.}\ \bibnamefont {Giapis}}, \ and\ \bibinfo
  {author} {\bibfnamefont {D.}~\bibnamefont {Poulikakos}},\ }\href {\doibase
  10.1103/PhysRevB.84.085442} {\bibfield  {journal} {\bibinfo  {journal} {Phys.
  Rev. B}\ }\textbf {\bibinfo {volume} {84}},\ \bibinfo {pages} {085442}
  (\bibinfo {year} {2011})}\BibitemShut {NoStop}%
\bibitem [{\citenamefont {Wong}\ \emph {et~al.}(2011)\citenamefont {Wong},
  \citenamefont {Léonard}, \citenamefont {Li},\ and\ \citenamefont
  {Wang}}]{Wong11}%
  \BibitemOpen
  \bibfield  {author} {\bibinfo {author} {\bibfnamefont {B.~M.}\ \bibnamefont
  {Wong}}, \bibinfo {author} {\bibfnamefont {F.}~\bibnamefont {Léonard}},
  \bibinfo {author} {\bibfnamefont {Q.}~\bibnamefont {Li}}, \ and\ \bibinfo
  {author} {\bibfnamefont {G.~T.}\ \bibnamefont {Wang}},\ }\href@noop {}
  {\bibfield  {journal} {\bibinfo  {journal} {Nano Letters}\ }\textbf {\bibinfo
  {volume} {11}},\ \bibinfo {pages} {3074} (\bibinfo {year}
  {2011})}\BibitemShut {NoStop}%
\bibitem [{\citenamefont {Bl{\"o}mers}\ \emph {et~al.}(2013)\citenamefont
  {Bl{\"o}mers}, \citenamefont {Rieger}, \citenamefont {Zellekens},
  \citenamefont {Haas}, \citenamefont {Lepsa}, \citenamefont {Hardtdegen},
  \citenamefont {G{\"u}l}, \citenamefont {Demarina}, \citenamefont
  {Gr{\"u}tzmacher}, \citenamefont {L{\"u}th},\ and\ \citenamefont
  {Sch{\"a}pers}}]{Blomers13}%
  \BibitemOpen
  \bibfield  {author} {\bibinfo {author} {\bibfnamefont {C.}~\bibnamefont
  {Bl{\"o}mers}}, \bibinfo {author} {\bibfnamefont {T.}~\bibnamefont {Rieger}},
  \bibinfo {author} {\bibfnamefont {P.}~\bibnamefont {Zellekens}}, \bibinfo
  {author} {\bibfnamefont {F.}~\bibnamefont {Haas}}, \bibinfo {author}
  {\bibfnamefont {M.~I.}\ \bibnamefont {Lepsa}}, \bibinfo {author}
  {\bibfnamefont {H.}~\bibnamefont {Hardtdegen}}, \bibinfo {author}
  {\bibfnamefont {{\"O}.}~\bibnamefont {G{\"u}l}}, \bibinfo {author}
  {\bibfnamefont {N.}~\bibnamefont {Demarina}}, \bibinfo {author}
  {\bibfnamefont {D.}~\bibnamefont {Gr{\"u}tzmacher}}, \bibinfo {author}
  {\bibfnamefont {H.}~\bibnamefont {L{\"u}th}}, \ and\ \bibinfo {author}
  {\bibfnamefont {T.}~\bibnamefont {Sch{\"a}pers}},\ }\href
  {http://stacks.iop.org/0957-4484/24/i=3/a=035203} {\bibfield  {journal}
  {\bibinfo  {journal} {Nanotechnology}\ }\textbf {\bibinfo {volume} {24}},\
  \bibinfo {pages} {035203} (\bibinfo {year} {2013})}\BibitemShut {NoStop}%
\bibitem [{\citenamefont {Qian}\ \emph {et~al.}(2012)\citenamefont {Qian},
  \citenamefont {Brewster}, \citenamefont {Lim}, \citenamefont {Ling},
  \citenamefont {Greene}, \citenamefont {Laboutin}, \citenamefont {Johnson},
  \citenamefont {Grade{\v{c}}ak}, \citenamefont {Cao},\ and\ \citenamefont
  {Li}}]{Qian12}%
  \BibitemOpen
  \bibfield  {author} {\bibinfo {author} {\bibfnamefont {F.}~\bibnamefont
  {Qian}}, \bibinfo {author} {\bibfnamefont {M.}~\bibnamefont {Brewster}},
  \bibinfo {author} {\bibfnamefont {S.~K.}\ \bibnamefont {Lim}}, \bibinfo
  {author} {\bibfnamefont {Y.}~\bibnamefont {Ling}}, \bibinfo {author}
  {\bibfnamefont {C.}~\bibnamefont {Greene}}, \bibinfo {author} {\bibfnamefont
  {O.}~\bibnamefont {Laboutin}}, \bibinfo {author} {\bibfnamefont {J.~W.}\
  \bibnamefont {Johnson}}, \bibinfo {author} {\bibfnamefont {S.}~\bibnamefont
  {Grade{\v{c}}ak}}, \bibinfo {author} {\bibfnamefont {Y.}~\bibnamefont {Cao}},
  \ and\ \bibinfo {author} {\bibfnamefont {Y.}~\bibnamefont {Li}},\ }\href@noop
  {} {\bibfield  {journal} {\bibinfo  {journal} {Nano Letters}\ }\textbf
  {\bibinfo {volume} {12}},\ \bibinfo {pages} {3344} (\bibinfo {year}
  {2012})}\BibitemShut {NoStop}%
\bibitem [{\citenamefont {Heurlin}\ \emph {et~al.}(2015)\citenamefont
  {Heurlin}, \citenamefont {Stankevi{\v{c}}}, \citenamefont
  {Mickevi{\v{c}}ius}, \citenamefont {Yngman}, \citenamefont {Lindgren},
  \citenamefont {Mikkelsen}, \citenamefont {Feidenhans’l}, \citenamefont
  {Borgst{\"o}m},\ and\ \citenamefont {Samuelson}}]{Heurlin15}%
  \BibitemOpen
  \bibfield  {author} {\bibinfo {author} {\bibfnamefont {M.}~\bibnamefont
  {Heurlin}}, \bibinfo {author} {\bibfnamefont {T.}~\bibnamefont
  {Stankevi{\v{c}}}}, \bibinfo {author} {\bibfnamefont {S.}~\bibnamefont
  {Mickevi{\v{c}}ius}}, \bibinfo {author} {\bibfnamefont {S.}~\bibnamefont
  {Yngman}}, \bibinfo {author} {\bibfnamefont {D.}~\bibnamefont {Lindgren}},
  \bibinfo {author} {\bibfnamefont {A.}~\bibnamefont {Mikkelsen}}, \bibinfo
  {author} {\bibfnamefont {R.}~\bibnamefont {Feidenhans’l}}, \bibinfo
  {author} {\bibfnamefont {M.~T.}\ \bibnamefont {Borgst{\"o}m}}, \ and\
  \bibinfo {author} {\bibfnamefont {L.}~\bibnamefont {Samuelson}},\ }\href@noop
  {} {\bibfield  {journal} {\bibinfo  {journal} {Nano Letters}\ }\textbf
  {\bibinfo {volume} {15}},\ \bibinfo {pages} {2462} (\bibinfo {year}
  {2015})}\BibitemShut {NoStop}%
\bibitem [{\citenamefont {Yuan}\ \emph {et~al.}(2015)\citenamefont {Yuan},
  \citenamefont {Caroff}, \citenamefont {Wang}, \citenamefont {Guo},
  \citenamefont {Wang}, \citenamefont {Jackson}, \citenamefont {Smith},
  \citenamefont {Tan},\ and\ \citenamefont {Jagadish}}]{Yuan15}%
  \BibitemOpen
  \bibfield  {author} {\bibinfo {author} {\bibfnamefont {X.}~\bibnamefont
  {Yuan}}, \bibinfo {author} {\bibfnamefont {P.}~\bibnamefont {Caroff}},
  \bibinfo {author} {\bibfnamefont {F.}~\bibnamefont {Wang}}, \bibinfo {author}
  {\bibfnamefont {Y.}~\bibnamefont {Guo}}, \bibinfo {author} {\bibfnamefont
  {Y.}~\bibnamefont {Wang}}, \bibinfo {author} {\bibfnamefont {H.~E.}\
  \bibnamefont {Jackson}}, \bibinfo {author} {\bibfnamefont {L.~M.}\
  \bibnamefont {Smith}}, \bibinfo {author} {\bibfnamefont {H.~H.}\ \bibnamefont
  {Tan}}, \ and\ \bibinfo {author} {\bibfnamefont {C.}~\bibnamefont
  {Jagadish}},\ }\href@noop {} {\bibfield  {journal} {\bibinfo  {journal}
  {Advanced Functional Materials}\ }\textbf {\bibinfo {volume} {25}},\ \bibinfo
  {pages} {5300} (\bibinfo {year} {2015})}\BibitemShut {NoStop}%
\bibitem [{\citenamefont {Sitek}\ \emph {et~al.}(2015)\citenamefont {Sitek},
  \citenamefont {Serra}, \citenamefont {Gudmundsson},\ and\ \citenamefont
  {Manolescu}}]{Sitek15}%
  \BibitemOpen
  \bibfield  {author} {\bibinfo {author} {\bibfnamefont {A.}~\bibnamefont
  {Sitek}}, \bibinfo {author} {\bibfnamefont {L.}~\bibnamefont {Serra}},
  \bibinfo {author} {\bibfnamefont {V.}~\bibnamefont {Gudmundsson}}, \ and\
  \bibinfo {author} {\bibfnamefont {A.}~\bibnamefont {Manolescu}},\ }\href
  {\doibase 10.1103/PhysRevB.91.235429} {\bibfield  {journal} {\bibinfo
  {journal} {Phys. Rev. B}\ }\textbf {\bibinfo {volume} {91}},\ \bibinfo
  {pages} {235429} (\bibinfo {year} {2015})}\BibitemShut {NoStop}%
\bibitem [{\citenamefont {P\"oyh\"onen}\ \emph {et~al.}(2014)\citenamefont
  {P\"oyh\"onen}, \citenamefont {Weststr\"om}, \citenamefont {R\"ontynen},\
  and\ \citenamefont {Ojanen}}]{Poyhonen14}%
  \BibitemOpen
  \bibfield  {author} {\bibinfo {author} {\bibfnamefont {K.}~\bibnamefont
  {P\"oyh\"onen}}, \bibinfo {author} {\bibfnamefont {A.}~\bibnamefont
  {Weststr\"om}}, \bibinfo {author} {\bibfnamefont {J.}~\bibnamefont
  {R\"ontynen}}, \ and\ \bibinfo {author} {\bibfnamefont {T.}~\bibnamefont
  {Ojanen}},\ }\href@noop {} {\bibfield  {journal} {\bibinfo  {journal} {Phys.
  Rev. B}\ }\textbf {\bibinfo {volume} {89}},\ \bibinfo {pages} {115109}
  (\bibinfo {year} {2014})}\BibitemShut {NoStop}%
\bibitem [{\citenamefont {Wakatsuki}\ \emph {et~al.}(2014)\citenamefont
  {Wakatsuki}, \citenamefont {Ezawa},\ and\ \citenamefont
  {Nagaosa}}]{Wakatsuki14}%
  \BibitemOpen
  \bibfield  {author} {\bibinfo {author} {\bibfnamefont {R.}~\bibnamefont
  {Wakatsuki}}, \bibinfo {author} {\bibfnamefont {M.}~\bibnamefont {Ezawa}}, \
  and\ \bibinfo {author} {\bibfnamefont {N.}~\bibnamefont {Nagaosa}},\
  }\href@noop {} {\bibfield  {journal} {\bibinfo  {journal} {Phys. Rev. B}\
  }\textbf {\bibinfo {volume} {89}},\ \bibinfo {pages} {174514} (\bibinfo
  {year} {2014})}\BibitemShut {NoStop}%
\bibitem [{\citenamefont {Sedlmayr}\ \emph {et~al.}(2016)\citenamefont
  {Sedlmayr}, \citenamefont {Aguiar-Hualde},\ and\ \citenamefont
  {Bena}}]{Sedlmayr16}%
  \BibitemOpen
  \bibfield  {author} {\bibinfo {author} {\bibfnamefont {N.}~\bibnamefont
  {Sedlmayr}}, \bibinfo {author} {\bibfnamefont {J.~M.}\ \bibnamefont
  {Aguiar-Hualde}}, \ and\ \bibinfo {author} {\bibfnamefont {C.}~\bibnamefont
  {Bena}},\ }\href {\doibase 10.1103/PhysRevB.93.155425} {\bibfield  {journal}
  {\bibinfo  {journal} {Phys. Rev. B}\ }\textbf {\bibinfo {volume} {93}},\
  \bibinfo {pages} {155425} (\bibinfo {year} {2016})}\BibitemShut {NoStop}%
\bibitem [{\citenamefont {Sitek}\ \emph {et~al.}(2016)\citenamefont {Sitek},
  \citenamefont {Thorgilsson}, \citenamefont {Gudmundsson},\ and\ \citenamefont
  {Manolescu}}]{Sitek16}%
  \BibitemOpen
  \bibfield  {author} {\bibinfo {author} {\bibfnamefont {A.}~\bibnamefont
  {Sitek}}, \bibinfo {author} {\bibfnamefont {G.}~\bibnamefont {Thorgilsson}},
  \bibinfo {author} {\bibfnamefont {V.}~\bibnamefont {Gudmundsson}}, \ and\
  \bibinfo {author} {\bibfnamefont {A.}~\bibnamefont {Manolescu}},\ }\href
  {http://stacks.iop.org/0957-4484/27/i=22/a=225202} {\bibfield  {journal}
  {\bibinfo  {journal} {Nanotechnology}\ }\textbf {\bibinfo {volume} {27}},\
  \bibinfo {pages} {225202} (\bibinfo {year} {2016})}\BibitemShut {NoStop}%
\bibitem [{\citenamefont {Winkler}(2004)}]{Winkler04}%
  \BibitemOpen
  \bibfield  {author} {\bibinfo {author} {\bibfnamefont {R.}~\bibnamefont
  {Winkler}},\ }\href@noop {} {\bibfield  {journal} {\bibinfo  {journal}
  {Physica E: Low-dimensional Systems and Nanostructures}\ }\textbf {\bibinfo
  {volume} {22}},\ \bibinfo {pages} {450 } (\bibinfo {year}
  {2004})}\BibitemShut {NoStop}%
\bibitem [{\citenamefont {Bringer}\ and\ \citenamefont
  {Sch\"apers}(2011)}]{Bringer11}%
  \BibitemOpen
  \bibfield  {author} {\bibinfo {author} {\bibfnamefont {A.}~\bibnamefont
  {Bringer}}\ and\ \bibinfo {author} {\bibfnamefont {T.}~\bibnamefont
  {Sch\"apers}},\ }\href@noop {} {\bibfield  {journal} {\bibinfo  {journal}
  {Phys. Rev. B}\ }\textbf {\bibinfo {volume} {83}},\ \bibinfo {pages} {115305}
  (\bibinfo {year} {2011})}\BibitemShut {NoStop}%
\bibitem [{\citenamefont {Manolescu}\ \emph {et~al.}(2013)\citenamefont
  {Manolescu}, \citenamefont {Rosdahl}, \citenamefont {Erlingsson},
  \citenamefont {Serra},\ and\ \citenamefont {Gudmundsson}}]{Manolescu13}%
  \BibitemOpen
  \bibfield  {author} {\bibinfo {author} {\bibfnamefont {A.}~\bibnamefont
  {Manolescu}}, \bibinfo {author} {\bibfnamefont {T.~O.}\ \bibnamefont
  {Rosdahl}}, \bibinfo {author} {\bibfnamefont {S.~I.}\ \bibnamefont
  {Erlingsson}}, \bibinfo {author} {\bibfnamefont {L.}~\bibnamefont {Serra}}, \
  and\ \bibinfo {author} {\bibfnamefont {V.}~\bibnamefont {Gudmundsson}},\
  }\href@noop {} {\bibfield  {journal} {\bibinfo  {journal} {The European
  Physical Journal B}\ }\textbf {\bibinfo {volume} {86}},\ \bibinfo {pages}
  {445} (\bibinfo {year} {2013})}\BibitemShut {NoStop}%
\bibitem [{\citenamefont {Lim}\ \emph {et~al.}(2013)\citenamefont {Lim},
  \citenamefont {Lopez},\ and\ \citenamefont {L.}}]{Lim13}%
  \BibitemOpen
  \bibfield  {author} {\bibinfo {author} {\bibfnamefont {J.~S.}\ \bibnamefont
  {Lim}}, \bibinfo {author} {\bibfnamefont {R.}~\bibnamefont {Lopez}}, \ and\
  \bibinfo {author} {\bibfnamefont {S.}~\bibnamefont {L.}},\ }\href@noop {}
  {\bibfield  {journal} {\bibinfo  {journal} {Europhys. Lett.}\ }\textbf
  {\bibinfo {volume} {103}},\ \bibinfo {pages} {37004} (\bibinfo {year}
  {2013})}\BibitemShut {NoStop}%
\bibitem [{\citenamefont {Serra}(2013)}]{ser13}%
  \BibitemOpen
  \bibfield  {author} {\bibinfo {author} {\bibfnamefont {L.}~\bibnamefont
  {Serra}},\ }\href {\doibase 10.1103/PhysRevB.87.075440} {\bibfield  {journal}
  {\bibinfo  {journal} {Phys. Rev. B}\ }\textbf {\bibinfo {volume} {87}},\
  \bibinfo {pages} {075440} (\bibinfo {year} {2013})}\BibitemShut {NoStop}%
\bibitem [{\citenamefont {Ghosh}\ \emph {et~al.}(2010)\citenamefont {Ghosh},
  \citenamefont {Sau}, \citenamefont {Tewari},\ and\ \citenamefont
  {Das~Sarma}}]{Ghosh2010}%
  \BibitemOpen
  \bibfield  {author} {\bibinfo {author} {\bibfnamefont {P.}~\bibnamefont
  {Ghosh}}, \bibinfo {author} {\bibfnamefont {J.~D.}\ \bibnamefont {Sau}},
  \bibinfo {author} {\bibfnamefont {S.}~\bibnamefont {Tewari}}, \ and\ \bibinfo
  {author} {\bibfnamefont {S.}~\bibnamefont {Das~Sarma}},\ }\href@noop {}
  {\bibfield  {journal} {\bibinfo  {journal} {Phys. Rev. B}\ }\textbf {\bibinfo
  {volume} {82}},\ \bibinfo {pages} {184525} (\bibinfo {year}
  {2010})}\BibitemShut {NoStop}%
\end{thebibliography}
%

\end{document}